\documentclass[pdflatex,sn-mathphys-ay]{sn-jnl}

\usepackage{graphicx}%
\usepackage{multirow}%
\usepackage{amsmath,amssymb,amsfonts}%
\usepackage{amsthm}%
\usepackage{mathrsfs}%
\usepackage{booktabs}
\usepackage{threeparttable}
\usepackage{subcaption}
\usepackage[ruled,vlined,linesnumbered]{algorithm2e}
\SetAlgoNlRelativeSize{-2}
\SetAlFnt{\scriptsize}
\SetAlCapFnt{\small}
\SetAlCapNameFnt{\small}
\newcommand{\E}{\mathbb{E}}
\usepackage{bm}

\usepackage{mathtools}
\usepackage[title]{appendix}%
\usepackage{xcolor}%
\usepackage{textcomp}%
\usepackage{manyfoot}%
\usepackage{booktabs}%
\usepackage{algpseudocode}%
\usepackage{listings}%
\hypersetup{
    colorlinks = true,
    citecolor = black,
    linkcolor = black,
    urlcolor  = black
}

\theoremstyle{thmstyleone}%
\theoremstyle{thmstyletwo}%

\theoremstyle{thmstylethree}%

\begin{document}

\title[Article Title]{Variational Inference for Functional Data Clustering via Dirichlet Process Mixtures with Correlated Errors}


\author*{\fnm{Chengqian} \sur{Xian}}\email{xianchengqian@hit.edu.cn}
\affil{\orgdiv{School of Science}, \orgname{Harbin Institute of Technology, Shenzhen}, \orgaddress{\postcode{518055}, \country{China}}}


\abstract{
We propose a Bayesian model-based approach for clustering functional data with an unknown number of clusters and within-curve correlated observations. Cluster-specific mean functions are represented using B-spline basis expansions, while within-curve dependence is modeled through an Ornstein--Uhlenbeck covariance structure. A truncated Dirichlet process mixture is used to infer the effective number of clusters, and a variational EM algorithm is developed for efficient posterior approximation. Simulation studies show that the proposed method performs well under both correctly specified and misspecified mean-function settings and achieves higher average values of the reported clustering metrics than the competing methods in the simulation settings considered. Comparisons with MCMC indicate that the variational approximation produces clustering results and parameter estimates that are in close agreement with those obtained by MCMC, while requiring substantially lower computational cost. An application to Canadian daily temperature curves further demonstrates the practical usefulness of the method in identifying interpretable functional clusters while accounting for within-curve dependence.
}




\maketitle

\section{Introduction}\label{IntroSec}

Functional data arise when the observational unit can naturally be viewed as a function or curve defined over a continuous domain, such as time, space, or wavelength \citep{RamsaySilverman2005}. Advances in data collection technologies have made such data increasingly common in fields including biomedical research, environmental science, engineering, energy systems, and neuroscience. In practice, the underlying functions are typically observed at a finite collection of evaluation points, and functional data analysis provides a framework for recovering and analyzing the continuous structures underlying these discrete measurements \citep{Gertheiss_2024}. Among the fundamental tasks in functional data analysis, clustering aims to identify latent groups of curves with similar underlying patterns when cluster memberships are unknown in advance.

A substantial literature has been developed for functional data clustering; see \cite{Jacques_2014} and, more recently, \cite{Zhang_2023} for comprehensive reviews. Existing approaches can broadly be viewed as dimension-reduction or filtering methods, distance-based methods, and model-based methods. Filtering approaches first represent the functions in a finite-dimensional space, for example through basis expansions or functional principal component analysis, and subsequently apply conventional clustering procedures to the resulting coefficients or scores. Distance-based approaches construct suitable dissimilarities between curves and apply algorithms such as functional $K$-means \citep{TarpeyKinateder2003} or hierarchical clustering. In contrast, model-based approaches introduce a probabilistic model for the heterogeneous functional population and infer the cluster structure jointly with the parameters governing the underlying curves \citep{Chamroukhi_2019, Xian_2025}. \cite{Centofanti_2024} proposed a sparse and smooth model-based clustering procedure that jointly identifies informative regions of the functional domain, while \cite{Anton_2024} considered mixtures of multivariate $t$ distributions to improve robustness against functional outliers. More recently, \cite{Rocci_2025} integrated dimension reduction and functional $K$-means within a single optimization framework, \cite{Chen_2025} developed a local clustering procedure that permits the cluster structure to vary over the functional domain, and \cite{Saeidi_2025} proposed a model-based clustering method based on probability-density approximation in a reproducing kernel Hilbert space.

An important issue in model-based functional clustering is the determination of the number of clusters. Conventional finite mixture models typically require the number of components to be specified in advance or selected using an additional model-selection criterion. Bayesian nonparametric mixture models provide an alternative in which the effective number of occupied clusters can be inferred from the data. The Dirichlet process (DP), in particular, provides a convenient probabilistic construction for this purpose through its stick-breaking representation \citep{Ferguson1973,Sethuraman1994,Ishwaran_2001}. DP-based models have previously been considered for clustering functional observations. For example, \citet{Park_2019} proposed a DP mixture of Fourier-series regression models for clustering temporal gene-expression profiles, with spike-and-slab priors used to identify important Fourier components. More recently, \citet{Gao_2024} developed a weighted Dirichlet process approach for clustering subject-specific functions represented through regression splines, together with a variable-selection mechanism for identifying important features of the functional trajectories. These studies demonstrate the usefulness of DP priors for functional clustering without requiring the number of clusters to be specified in advance. However, their primary focus is on flexible representation and clustering of the mean functions via Markov chain Monte Carlo (MCMC), rather than on explicitly modeling structured within-curve error dependence together with computationally efficient Bayesian inference.

A second important issue, particularly for densely observed functional data, concerns dependence among observations from the same curve. Measurements taken at nearby evaluation points are typically correlated, and the strength of this dependence may decrease as the distance between evaluation points increases. Nevertheless, simplified error structures are used to facilitate computation. Failure to accommodate such dependence can distort the characterization of residual variability and may consequently affect the inferred clustering structure. Correlation modeling has long been recognized as important in functional data analysis; for example, \citet{Dias_2013} considered correlated error structures in a hierarchical functional model. More recent work has further emphasized the importance of accounting for within-curve dependence. \citet{daCruz_2024} developed a Bayesian variational approach for functional representation that explicitly accommodates correlated errors within curves. \citet{Iwashige_2025} showed that misspecification of the error structure in Bayesian functional clustering can lead to overestimation of the number of clusters and demonstrated substantial improvements when the underlying correlation structure is properly incorporated.

Accounting simultaneously for an unknown number of clusters and within-curve correlation leads to a flexible but computationally challenging model. Posterior inference for Dirichlet process mixtures is traditionally carried out using MCMC, but posterior sampling can become computationally expensive when the number of curves or the number of observations per curve is large. Variational inference (VI) offers an alternative by replacing posterior sampling with deterministic optimization; see \cite{David_2017} for a comprehensive review. In functional data clustering settings, \cite{Xian_2025} proposed a variational Bayes algorithm for simultaneous clustering and smoothing based on a finite B-spline regression mixture model. The approach by \cite{Xian_2025} provides efficient posterior approximation but determines the number of clusters separately through the deviance information criterion \citep{SpiegelhalterEtAl2002} and does not explicitly model serial correlation in the residual process.

Motivated by these considerations, we develop a Bayesian approach for clustering functional data that combines a truncated Dirichlet process mixture with an explicit within-curve correlation structure. Conditional on cluster membership, the cluster-specific mean function is represented using a B-spline basis expansion. Rather than assuming conditionally independent errors, we introduce an Ornstein--Uhlenbeck (OU) covariance structure \citep{UhlenbeckOrnstein_1930}, under which correlation decreases exponentially with the distance between evaluation points. The DP prior provides a data-adaptive mechanism for determining the effective number of clusters, while the OU covariance offers a parsimonious and interpretable representation of within-curve dependence.


For posterior computation, we develop a variational EM algorithm in which the variational E-step updates the cluster assignments, stick-breaking variables, B-spline coefficients, and cluster-specific precision parameters using coordinate ascent variational inference, exploiting their conditional conjugacy. Since the OU decay parameter enters the covariance matrix nonlinearly and does not yield a conjugate variational update, it is treated as a deterministic parameter and updated in the M-step by maximizing the evidence lower bound. This construction preserves closed-form coordinate updates for the conjugate components while allowing the strength of within-curve dependence to be estimated from the data. To evaluate the accuracy and computational advantages of the variational approximation, we additionally develop an MCMC sampler under the same probabilistic model, using Gibbs updates for the conjugate components and a Metropolis--Hastings step for the OU decay parameter.

Simulation studies investigate the proposed approach under different strengths of within-curve correlation and different latent clustering structures. Particular attention is given to clustering accuracy, estimation of the correlation parameter, and comparison between variational and MCMC posterior inference. We also compare the proposed procedure with alternative functional clustering approaches and illustrate its practical performance using real functional data. Our proposed method is implemented in R, and the code is available at \url{https://github.com/chengqianxian/funclustDPOU}.

The remainder of this paper is organized as follows. Section~\ref{MethodSec} introduces the proposed Dirichlet process mixture model with correlated errors and develops the variational EM and MCMC estimation procedures. Section~\ref{SimSec} presents the simulation studies. Section~\ref{AppSec} applies the proposed methodology to real data. Section~\ref{ConDisSec} concludes the paper with a discussion and directions for future research.

\section{Methodology}\label{MethodSec}

\subsection{Model Specification}

Let $Y_i = (Y_i(t_{i1}), \dots, Y_i(t_{in_i}))^\top \in \mathbb{R}^{n_i}$ denote the observed data vector for the $i$-th curve, measured at evaluation points $t_i = (t_{i1}, \dots, t_{in_i})^\top$, for $i = 1, \dots, N$. We assume that each curve arises from an unknown latent cluster indexed by $c_i \in \{1, \dots, H\}$, where $H$ is a truncation level. Conditional on cluster membership $c_i = h$, the cluster-specific mean function, $f_h(t)$, is modeled using a B-spline basis expansion
\[
f_h(t) = \sum_{m=1}^M B_m(t)\,\phi_{hm},
\]
where $\{B_m(\cdot)\}_{m=1}^M$ are predefined basis functions and $\phi_h = (\phi_{h1}, \dots, \phi_{hM})^\top$ is the coefficient vector for cluster $h$. Let $B_i$ denote the $n_i \times M$ design matrix with entries $(B_i)_{jm} = B_m(t_{ij})$, so that the mean vector is given by $B_i \phi_h$. As in \cite{Xian_2025}, we employ cubic B-splines with equally spaced knots and assume that the number of basis functions $M$ is predefined and fixed. 

To model within-curve dependence, we assume that the residual process follows a Gaussian process with an Ornstein--Uhlenbeck (OU) covariance structure \citep{Dias_2013}. Specifically, conditional on $c_i = h$, we assume
\[
Y_i \mid \{c_i = h, \phi_h, \tau_h, \delta\}
\sim
\mathcal{N}\bigl( B_i \phi_h,\ \tau_h^{-1} \Omega_i(\delta) \bigr),
\]
where $\tau_h = 1/\sigma_h^2$ is a cluster-specific precision parameter, and the covariance matrix $\Omega_i(\delta)$ is defined element-wise as
\[
\bigl(\Omega_i(\delta)\bigr)_{ks}
=
\exp\bigl(-\delta |t_{ik} - t_{is}|\bigr),
\qquad \delta > 0.
\]
The parameter $\delta$ controls the rate of correlation decay and thus governs the strength of within-curve dependence.

To allow the number of clusters to be inferred from the data, we adopt a Dirichlet process mixture model using a truncated stick-breaking representation \citep{Ishwaran_2001}. The mixture weights, $\pi_1,\ldots,\pi_H$, are constructed as
\[
v_h \sim \mathrm{Beta}(1, \alpha), \;
\pi_h = v_h \prod_{\ell < h} (1 - v_\ell), \; h = 1, \dots, H,\,
\]
where $v_H=1$ such that $\sum_{h=1}^H \pi_h=1$ and $\alpha > 0$ is the concentration parameter. The cluster assignments are then given by
\[
c_i \mid \{\pi_h\}_{h=1}^H \sim \mathrm{Categorical}(\pi_1, \dots, \pi_H).
\]
It is useful to distinguish between the stick-breaking variables
$\boldsymbol v=(v_1,\ldots,v_{H-1})$ and the resulting mixture weights
$\boldsymbol\pi=(\pi_1,\ldots,\pi_H)$. The mixture weights are not an
additional independent set of model parameters; rather, they are deterministic
functions of the stick-breaking variables. The proposed model extends the work by \cite{Xian_2025} in two directions. First, the Dirichlet process prior enables data-driven determination of the number of clusters. Second, the OU covariance structure captures within-curve dependence within each curve, providing a more flexible alternative to independent error assumptions.


\subsection{Estimation via variational EM algorithm}
We consider a Bayesian framework and perform parameter estimation using variational inference. The exact posterior is analytically intractable due to the mixture structure and the OU-induced within-curve dependence, which yields likelihood terms involving covariance matrix inverses and log-determinants that depend nonlinearly on $\delta$. To address this, we develop a variational EM algorithm \citep{Coviello_2012, El_2016}, which provides a computationally efficient approximation to the posterior distribution. Specifically, we approximate the joint posterior of the latent variables and model parameters by a tractable variational distribution, while treating $\delta$ as a deterministic parameter to be optimized via the evidence lower bound (ELBO). 

Apart from the Dirichlet process prior on the mixture weights, we specify conjugate priors for the cluster-specific parameters. In particular, for each cluster $h = 1, \dots, H$, we assume
\[
\phi_h \sim \mathcal{N}(m_0, S_0),
\qquad
\tau_h \sim \mathrm{Gamma}(a_0, b_0),
\]
where the Gamma distribution is parameterized in terms of shape and rate. The hyperparameters $m_0 \in \mathbb{R}^M$, $S_0 \in \mathbb{R}^{M \times M}$, and $a_0, b_0 > 0$ are assumed to be fixed and known. 

We describe the proposed variational EM algorithm to infer the model parameters and latent variables, which is summarized in Algorithm \ref{alg:DPFunctionalOU}. Let $
\Theta=\{\boldsymbol c,\boldsymbol v,\boldsymbol\phi,\boldsymbol\tau\}$
denote the collection of latent variables and model parameters to which
variational distributions are assigned, excluding the deterministic covariance parameter
$\delta$. Here, the mixture weights $\boldsymbol\pi$ are not included as a
separate parameter block because they are deterministic functions of the
stick-breaking variables $\boldsymbol v$. Therefore, the variational
distribution assigned to $\boldsymbol v$ implicitly induces a distribution for
$\boldsymbol\pi$. Given $\delta$ and the data, the posterior distribution is $p(\Theta \mid Y, \delta)$. We approximate the posterior using a mean-field variational distribution
\[
q(\Theta)
=
\prod_{i=1}^N q(c_i)
\prod_{h=1}^{H-1} q(v_h)
\prod_{h=1}^H q(\phi_h)
\prod_{h=1}^H q(\tau_h).
\]

\subsubsection{Variational E-step}
The variational E-step proceeds by maximizing the ELBO defined as
\[
\mathcal{L}(q, \delta)
=
\mathbb{E}_q \big[ \log p(Y, \Theta \mid \delta) \big]
-
\mathbb{E}_q \big[ \log q(\Theta) \big],
\] 
which is equivalent to minimizing the Kullback--Leibler (KL) divergence \citep{KullbackLeibler1951} between the variational distribution $q(\Theta)$ and the true posterior distribution $p(\Theta \mid Y, \delta)$. Given a fixed value of $\delta$, we update the variational factors $q(c_i)$, $q(v_h)$, $q(\phi_h)$, and $q(\tau_h)$ using coordinate ascent variational inference (CAVI) under the mean-field assumption \citep{Bishop_2006}. In particular, the update equation for each variational factor can be obtained by taking the expectation of the so-called complete-data log-likelihood, $\log p(Y, \Theta \mid \delta)$, over the variational distribution of all random variables except the one of interest. For example, the variational update for $q(\phi_h)$ is obtained from
\[
\log q(\phi_h)
=
\mathbb{E}_{-\phi_h}
\left[
\log p(Y,\Theta\mid\delta)
\right]
+\text{const},
\]
where $\mathbb{E}_{-\phi_h}(\cdot)$ denotes expectation with respect to the variational distributions of all random variables except $\phi_h$. 

The complete-data log-likelihood can be written as
\begin{align}
\log p(\boldsymbol Y,\Theta\mid\delta)
={}&
\sum_{i=1}^N
\log p(Y_i\mid c_i,\boldsymbol\phi,\boldsymbol\tau,\delta)
+
\log p\{\boldsymbol c\mid\boldsymbol\pi(\boldsymbol v)\}
\nonumber\\
&+
\sum_{h=1}^{H-1}\log p(v_h)
+
\sum_{h=1}^{H}\log p(\boldsymbol\phi_h)
+
\sum_{h=1}^{H}\log p(\tau_h).
\label{eq:complete_loglik}
\end{align}
For notational simplicity, we subsequently write
$p(\boldsymbol c\mid\boldsymbol v)$ for
$p\{\boldsymbol c\mid\boldsymbol\pi(\boldsymbol v)\}$.

In what follows, we present the update equation for each variational factor with the derivation details provided in Appendix \ref{secA1}. All expectations in the update equations are derived and given in Section \ref{ExpSubsec}.

\paragraph{(i). Update for $q(\phi_h)$}

The variational posterior for the cluster-specific spline coefficients $\phi_h$ follows a multivariate normal distribution:
\[
q(\phi_h) = \mathcal{N}(\mu_h, \Sigma_h),
\]
with
\[
\Sigma_h^{-1}
=
S_0^{-1}
+
\mathbb{E}_q[\tau_h]
\sum_{i=1}^N r_{ih} B_i^\top \Omega_i(\delta)^{-1} B_i,
\]
\[
\mu_h
=
\Sigma_h
\left[
S_0^{-1} m_0
+
\mathbb{E}_q[\tau_h]
\sum_{i=1}^N r_{ih} B_i^\top \Omega_i(\delta)^{-1} Y_i
\right],
\]
where $r_{ih}$ is defined in Equation (\ref{r_ih}).

\paragraph{(ii). Update for $q(\tau_h)$}
\[
q(\tau_h) = \mathrm{Gamma}(\tilde{a}_h, \tilde{b}_h),
\]
with
\[
\tilde{a}_h = a_0 + \frac{1}{2} \sum_{i=1}^N r_{ih} n_i,
\]
\[
\tilde{b}_h = b_0 + \frac{1}{2} \sum_{i=1}^N r_{ih}
\mathbb{E}_q[(Y_i - B_i \phi_h)^\top \Omega_i^{-1} (Y_i - B_i \phi_h)].
\]

\paragraph{(iii). Update for $q(c_i)$}
\[
q(c_i)=\mathrm{Categorical}(r_{i1},\dots,r_{iH}),
\]
with
\begin{equation}\label{r_ih}
    r_{ih}
=
\frac{\exp(\alpha_{ih})}{\sum_{m=1}^H \exp(\alpha_{im})},
\end{equation}
where
\[
\alpha_{ih}
=
\mathbb{E}_q[\log \pi_h]
+
\frac{n_i}{2} \mathbb{E}_q[\log \tau_h]
-
\frac{1}{2}\log |\Omega_i(\delta)|
-
\frac{1}{2}\mathbb{E}_q[\tau_h] \mathbb{E}_q
\big[
(Y_i - B_i \phi_h)^\top \Omega_i^{-1} (Y_i - B_i \phi_h)
\big].
\]
Although $\pi_h$ does not have a separate variational factor, the quantity
$E_q[\log\pi_h]$ is obtained directly from the variational distributions of the
stick-breaking variables. For $h=1,\ldots,H-1$,
\begin{equation}
\log \pi_h
=
\log v_h
+
\sum_{\ell<h}\log(1-v_\ell),\nonumber
\end{equation}
and therefore
\begin{equation}
\mathbb{E}_q[\log\pi_h]
=
\mathbb{E}_q[\log v_h]
+
\sum_{\ell<h}\mathbb{E}_q[\log(1-v_\ell)].\nonumber
\end{equation}
For the final component, since $v_H=1$,
\begin{equation}
\pi_H
=
\prod_{\ell=1}^{H-1}(1-v_\ell),\nonumber
\end{equation}
so that
\begin{equation}
\mathbb{E}_q[\log\pi_H]
=
\sum_{\ell=1}^{H-1}\mathbb{E}_q[\log(1-v_\ell)].\nonumber
\end{equation}
The required expectations are given explicitly in
Section~\ref{ExpSubsec}.
\paragraph{(iv). Update for $q(v_h)$}
For $h=1,\ldots,H-1$, the variational posterior of the stick-breaking variable
$v_h$ is
\[
q(v_h) = \mathrm{Beta}(\gamma_{h1}, \gamma_{h2}),
\]
with
\[
\gamma_{h1} = 1 + \sum_{i=1}^N r_{ih}, \qquad
\gamma_{h2} = \alpha + \sum_{i=1}^N \sum_{\ell > h} r_{i\ell}.
\]

\subsubsection{M-step}
In the M-step, we update the decay $\delta$ in the OU covariance by further maximizing the ELBO with respect to $\delta$:
\[
\hat{\delta}
=
\arg\max_{\delta > 0} \mathcal{L}(q, \delta).
\]
Note that
\[
\mathcal{L}(q, \delta)
=
\mathbb{E}_q \big[ \log p(Y, \Theta \mid \delta) \big]
-
\mathbb{E}_q \big[ \log q(\Theta) \big],
\]
where $\log p(Y, \Theta \mid \delta)$ can be decomposed as in Equation (\ref{eq:complete_loglik}). For the log-likelihood term, conditional on $c_i = h$, we have
\begin{align*}
\log p(Y_i \mid c_i = h, \phi_h, \tau_h, \delta)
&=
-\frac{n_i}{2}\log(2\pi)
+\frac{n_i}{2}\log \tau_h
\\
&\quad
-\frac{1}{2}\log |\Omega_i(\delta)|
-\frac{\tau_h}{2}
(Y_i - B_i \phi_h)^\top \Omega_i(\delta)^{-1} (Y_i - B_i \phi_h).
\end{align*}

Taking expectation with respect to $q$, the ELBO becomes
\begin{align*}
\mathcal{L}(q, \delta)
&=
\sum_{i=1}^N \sum_{h=1}^H r_{ih}
\Bigg[
\frac{n_i}{2} \mathbb{E}_q[\log \tau_h]
-
\frac{1}{2}\log |\Omega_i(\delta)|
\\
&\quad
-
\frac{1}{2} \mathbb{E}_q[\tau_h]
\mathbb{E}_q
\big[
(Y_i - B_i \phi_h)^\top \Omega_i^{-1} (Y_i - B_i \phi_h)
\big]
\Bigg]
\\
&\quad
+ \mathbb{E}_q[\log p(c \mid v)]
+ \sum_{h=1}^{H-1} \mathbb{E}_q[\log p(v_h)]
+ \sum_{h=1}^H \mathbb{E}_q[\log p(\phi_h)]
+ \sum_{h=1}^H \mathbb{E}_q[\log p(\tau_h)]
\\
&\quad
-
\sum_{i=1}^N \mathbb{E}_q[\log q(c_i)]
-
\sum_{h=1}^{H-1} \mathbb{E}_q[\log q(v_h)]
-
\sum_{h=1}^H \mathbb{E}_q[\log q(\phi_h)]
-
\sum_{h=1}^H \mathbb{E}_q[\log q(\tau_h)].
\end{align*}

The expectation terms in the ELBO are provided in Section \ref{ExpSubsec}. In the M-step, \(q\) is fixed. Therefore, all variational parameters are treated as constants. Only \(\delta\) changes through the OU covariance matrix
\[
\Omega_i(\delta)_{ks}
=
\exp\{-\delta |t_{ik}-t_{is}|\}.
\]

From the ELBO, the only \(\delta\)-dependent part is the expected log-likelihood:
\[
\mathcal{L}_{\delta}
=
\sum_{i=1}^{N}
\sum_{h=1}^{H}
r_{ih}
\left[
-\frac{1}{2}\log|\Omega_i(\delta)|
-
\frac{1}{2}E_q[\tau_h]Q_{ih}(\delta)
\right],
\]
where
\[
Q_{ih}(\delta)
=
\E_q\left[
(Y_i-B_i\phi_h)^\top
\Omega_i(\delta)^{-1}
(Y_i-B_i\phi_h)
\right].
\]

Therefore, the M-step is
\[
\hat{\delta}
=
\arg\max_{\delta>0}
\sum_{i=1}^{N}
\sum_{h=1}^{H}
r_{ih}
\left[
-\frac{1}{2}\log|\Omega_i(\delta)|
-
\frac{1}{2}E_q[\tau_h]Q_{ih}(\delta)
\right].
\]

\subsubsection{Expectations in the ELBO and update equations}\label{ExpSubsec}

The coordinate ascent updates and the ELBO involve several expectations with respect to the variational distributions. In this subsection, we derive and summarize their closed-form expressions.

\paragraph{(i). Expectations involving $\tau_h$}
Since $q(\tau_h) = \mathrm{Gamma}(\tilde{a}_h, \tilde{b}_h)$, we have
\[
\mathbb{E}_q[\tau_h] = \frac{\tilde{a}_h}{\tilde{b}_h},
\qquad
\mathbb{E}_q[\log \tau_h] = \psi(\tilde{a}_h) - \log \tilde{b}_h,
\]
where $\psi(\cdot)$ denotes the digamma function.

\paragraph{(ii). Quadratic expectation}
For $q(\phi_h) = \mathcal{N}(\mu_h, \Sigma_h)$, the quadratic term appearing in the updates is given by
\begin{align*}
&\mathbb{E}_q
\big[
(Y_i - B_i \phi_h)^\top \Omega_i(\delta)^{-1} (Y_i - B_i \phi_h)
\big]
\\
&=
(Y_i - B_i \mu_h)^\top \Omega_i(\delta)^{-1} (Y_i - B_i \mu_h)
+
\mathrm{tr}\big( B_i^\top \Omega_i(\delta)^{-1} B_i \Sigma_h \big),
\end{align*}
using the fact that 
$\mathbb{E}(X^\top A X)
=
\mathbb{E}(X)^\top A\,\mathbb{E}(X)
+
\mathrm{tr}\!\left(A\,\mathrm{Var}(X)\right),
$ for a random vector $X$ and a symmetric matrix $A$.

\paragraph{(iii). Expectations involving the stick-breaking weights}

Under the variational distribution
$q(v_h) = \mathrm{Beta}(\gamma_{h1}, \gamma_{h2})$,
for $h=1,\ldots,H-1$, the expectations required for updating $q(c_i)$ are
\[
\mathbb{E}_q[\log v_h]
=
\psi(\gamma_{h1}) - \psi(\gamma_{h1} + \gamma_{h2}),
\]
\[
\mathbb{E}_q[\log(1 - v_h)]
=
\psi(\gamma_{h2}) - \psi(\gamma_{h1} + \gamma_{h2}).
\]

Using the stick-breaking representation, for $h=1,\ldots,H-1$,
\[
\pi_h = v_h \prod_{\ell < h} (1 - v_\ell),
\]
the expectation of $\log \pi_h$ is
\begin{align*}
\mathbb{E}_q[\log \pi_h]
&=
\mathbb{E}_q[\log v_h]
+
\sum_{\ell<h}
\mathbb{E}_q[\log(1-v_\ell)]
\\
&=
\psi(\gamma_{h1})-\psi(\gamma_{h1}+\gamma_{h2})
+
\sum_{\ell<h}
\left[
\psi(\gamma_{\ell2})
-
\psi(\gamma_{\ell1}+\gamma_{\ell2})
\right].
\end{align*}

For the final component, $v_H=1$, and hence
\[
\pi_H
=
\prod_{\ell=1}^{H-1}(1-v_\ell).
\]
Therefore,
\begin{align*}
\mathbb{E}_q[\log \pi_H]
&=
\sum_{\ell=1}^{H-1}
\mathbb{E}_q[\log(1-v_\ell)]
\\
&=
\sum_{\ell=1}^{H-1}
\left[
\psi(\gamma_{\ell2})
-
\psi(\gamma_{\ell1}+\gamma_{\ell2})
\right].
\end{align*}

\paragraph{(iv). Prior and allocation expectations}
The expectations of the prior and allocation terms are given as follows.
\begin{align*}
\mathbb{E}_q[\log p(\phi_h)]
&=
-\frac{M}{2}\log(2\pi)
-\frac{1}{2}\log |S_0|
\\
&\quad
-\frac{1}{2}
\left[
\mathrm{tr}(S_0^{-1}\Sigma_h)
+
(\mu_h-m_0)^\top S_0^{-1}(\mu_h-m_0)
\right].
\end{align*}
For the allocation model,
\[
\log p(c\mid v)
=
\sum_{i=1}^N\sum_{h=1}^H \mathbf{1}(c_i=h)\log \pi_h,
\]
and hence
\[
\mathbb{E}_q[\log p(c\mid v)]
=
\sum_{i=1}^N\sum_{h=1}^H
r_{ih}\mathbb{E}_q[\log \pi_h].
\]
Lastly,
\[
\mathbb{E}_q[\log p(\tau_h)]
=
a_0\log b_0
-\log\Gamma(a_0)
+
(a_0-1)\mathbb{E}_q[\log\tau_h]
-
b_0\mathbb{E}_q[\tau_h].
\]
For $h=1,\ldots,H-1$,
\[
\mathbb{E}_q[\log p(v_h)]
=
-\log\alpha
+
(\alpha-1)\mathbb{E}_q[\log(1-v_h)].
\]

\paragraph{(v). Entropy terms}
The entropy terms in the ELBO are obtained directly from the variational factors.
\[
\mathbb{E}_q[\log q(\phi_h)]=
-\frac{M}{2}\log(2\pi)
-\frac{1}{2}\log|\Sigma_h|
-\frac{M}{2},
\]
\[
\mathbb{E}_q[\log q(\tau_h)]=
\tilde{a}_h\log \tilde{b}_h
-\log \Gamma(\tilde{a}_h)
+(\tilde{a}_h-1)\mathbb{E}_q[\log \tau_h]
-\tilde{b}_h\mathbb{E}_q[\tau_h],
\]
\[\mathbb{E}_q[\log q(c_i)]=
\sum_{h=1}^H r_{ih}\log r_{ih}.\]
For $h=1,\ldots,H-1$,
\[\mathbb{E}_q[\log q(v_h)]=
-\log B(\gamma_{h1},\gamma_{h2})
+(\gamma_{h1}-1)\mathbb{E}_q[\log v_h]
+(\gamma_{h2}-1)\mathbb{E}_q[\log(1-v_h)],
\]
where $B(\cdot,\cdot)$ denotes the Beta function.

\subsubsection{Initialization}
The variational EM algorithm (Algorithm \ref{alg:DPFunctionalOU}) requires initial values for the variational factors associated with the cluster assignments, stick-breaking weights, precision parameters, and the OU decay parameter. In our implementation, only the quantities required before the first coordinate update are initialized explicitly. The variational parameters for the spline coefficients, $\mu_h$ and $\Sigma_h$, are not numerically initialized because they are updated at the beginning of each iteration before being used in the remaining updates.

Specifically, the OU decay parameter is initialized as a positive constant, denoted by $\delta^{(0)}$ (for example, $\delta^{(0)} = 2$). The responsibility matrix $r^{(0)}=(r_{ih}^{(0)})$ is initialized as an $N\times H$ matrix whose rows sum to one. This can be done either by random initialization or by using a preliminary clustering method such as $K$-means. For example, if $K$-means is used with $K_0$ initial groups, the responsibility matrix can be initialized by assigning high probability to the corresponding $K$-means cluster and a small positive probability to the remaining mixture components, followed by row normalization.

Given the initial responsibility matrix, the variational parameters for the stick-breaking weights are initialized by
\[
\gamma_{h1}^{(0)}
=
1+\sum_{i=1}^N r_{ih}^{(0)},
\]
and
\[
\gamma_{h2}^{(0)}
=
\alpha+\sum_{i=1}^N\sum_{\ell>h}r_{i\ell}^{(0)},
\qquad h=1,\ldots,H-1,
\]
which are consistent with the coordinate ascent updates.

The variational precision parameters are initialized using a balanced-cluster approximation. Since the update of $\tilde a_h$ depends on the effective number of observations assigned to cluster $h$, we initialize it by assuming that the $N$ curves are approximately evenly distributed among the $H$ truncated components, giving
\[
\tilde a_h^{(0)}
=
a_0+\frac{1}{2}\bar n\frac{N}{H},
\]
where $\bar n=N^{-1}\sum_{i=1}^N n_i$ is the average number of observations per curve. The rate parameter is initialized as
\[
\tilde b_h^{(0)}=b_0+1,
\]
which avoids using the quadratic residual term before the spline coefficient posterior has been updated, while still yielding a finite and stable initial value for $\mathbb{E}_q[\tau_h]$.

\subsection{Estimation via MCMC algorithm}

To assess the accuracy of the proposed variational approximation, we additionally develop an MCMC sampler under the same truncated Dirichlet process mixture model. The sampler combines Gibbs updates for the conjugate model components, including the cluster-specific spline coefficients, precision parameters, cluster allocations, and stick-breaking weights, with a Metropolis--Hastings update for the OU decay parameter $\delta$. Posterior inference is then based on samples collected after discarding an initial burn-in period. Initialization of the MCMC sampler follows the same strategy as that adopted for the variational EM algorithm. The resulting MCMC procedure is summarized in Algorithm~\ref{alg:MCMC_DPFunctionalOU}. The Gaussian and Gamma full conditional parameters used in Algorithm~\ref{alg:MCMC_DPFunctionalOU} are
\[
\Sigma_h^\star
=
\left[
S_0^{-1}
+
\tau_h
\sum_{i:c_i=h}
B_i^\top\Omega_i(\delta)^{-1}B_i
\right]^{-1},
\]
\[
\mu_h^\star
=
\Sigma_h^\star
\left[
S_0^{-1}m_0
+
\tau_h
\sum_{i:c_i=h}
B_i^\top\Omega_i(\delta)^{-1}Y_i
\right],
\]
\[
a_h^\star
=
a_0+\frac12\sum_{i:c_i=h}n_i,
\qquad
b_h^\star
=
b_0+
\frac12\sum_{i:c_i=h}
(Y_i-B_i\phi_h)^\top
\Omega_i(\delta)^{-1}
(Y_i-B_i\phi_h).
\]

\begin{algorithm}[!ht]
\scriptsize
\DontPrintSemicolon

\KwData{ Observed functional data $\{Y_i, t_i, B_i\}_{i=1}^N$, where $Y_i=(Y_i(t_{i1}),\ldots,Y_i(t_{in_i}))^\top$, $t_i=(t_{i1},\ldots,t_{in_i})^\top$, and $B_i$ is the corresponding B-spline design matrix; truncation level $H$; hyperparameters $m_0$, $S_0$, $a_0$, $b_0$, and $\alpha$; convergence threshold $\gamma$ and maximum number of iterations $M$. } 

\KwResult{ Variational posterior distributions $q(c_i)$, $q(v_h)$, $q(\phi_h)$, and $q(\tau_h)$, with parameters $r_{ih}$, $\gamma_{h1}$, $\gamma_{h2}$, $\mu_h$, $\Sigma_h$, $\tilde a_h$, and $\tilde b_h$, together with the estimated OU decay parameter $\hat\delta$. } 

\textbf{Initialization}: initialize $r_{ih}^{(0)}$, $\tilde a_h^{(0)}$, $\tilde b_h^{(0)}$, $\gamma_{h1}^{(0)}$, $\gamma_{h2}^{(0)}$, and $\delta^{(0)}$; set $m=0$ and $\mathrm{ELBO}^{(0)}=-\infty$\;

\While{$m<M$ and ELBO difference $>\gamma$}{
$m\leftarrow m+1$\;

Construct
$\Omega_i^{(m-1)}=\Omega_i(\delta^{(m-1)})$ with
$\{\Omega_i^{(m-1)}\}_{rs}=\exp\{-\delta^{(m-1)}|t_{ir}-t_{is}|\}$\;

Calculate
$\E_q[\tau_h]=\tilde a_h^{(m-1)}/\tilde b_h^{(m-1)}$,
$\E_q[\log\tau_h]=\psi(\tilde a_h^{(m-1)})-\log \tilde b_h^{(m-1)}$, and
$\E_q[\log\pi_h]$\;

\For{$h=1,\ldots,H$}{
Update $q(\phi_h)=N(\mu_h,\Sigma_h)$:
\[
\begin{aligned}
\Sigma_h^{(m)}
&\leftarrow
\left[
S_0^{-1}
+
\E_q[\tau_h]
\sum_{i=1}^N
r_{ih}^{(m-1)}
B_i^\top
(\Omega_i^{(m-1)})^{-1}
B_i
\right]^{-1},
\\[-1mm]
\mu_h^{(m)}
&\leftarrow
\Sigma_h^{(m)}
\left[
S_0^{-1}m_0
+
\E_q[\tau_h]
\sum_{i=1}^N
r_{ih}^{(m-1)}
B_i^\top
(\Omega_i^{(m-1)})^{-1}
Y_i
\right].
\end{aligned}
\]
}

Calculate
\[
Q_{ih}^{(m)}
=
(Y_i-B_i\mu_h^{(m)})^\top
(\Omega_i^{(m-1)})^{-1}
(Y_i-B_i\mu_h^{(m)})
+
\operatorname{tr}
\{B_i^\top(\Omega_i^{(m-1)})^{-1}B_i\Sigma_h^{(m)}\}.
\]

\For{$h=1,\ldots,H$}{
Update $q(\tau_h)=\mathrm{Gamma}(\tilde a_h,\tilde b_h)$:
\[
\tilde a_h^{(m)}
\leftarrow
a_0+\frac12\sum_{i=1}^N r_{ih}^{(m-1)}n_i,
\qquad
\tilde b_h^{(m)}
\leftarrow
b_0+\frac12\sum_{i=1}^N r_{ih}^{(m-1)}Q_{ih}^{(m)}.
\]
}

\For{$h=1,\ldots,H-1$}{
Update $q(v_h)=\mathrm{Beta}(\gamma_{h1},\gamma_{h2})$:
\[
\gamma_{h1}^{(m)}
\leftarrow
1+\sum_{i=1}^N r_{ih}^{(m-1)},
\qquad
\gamma_{h2}^{(m)}
\leftarrow
\alpha+\sum_{i=1}^N\sum_{\ell>h}r_{i\ell}^{(m-1)}.
\]
Set $v_H=1$ implicitly when evaluating the mixture weights
$\pi_1,\ldots,\pi_H$.
}

Recalculate $\E_q[\tau_h]$, $\E_q[\log\tau_h]$, and $\E_q[\log\pi_h]$\;

\For{$i=1,\ldots,N$}{
Calculate
\[
A_{ih}^{(m)}
=
\E_q[\log\pi_h]
+
\frac{n_i}{2}\E_q[\log\tau_h]
-
\frac12\log|\Omega_i^{(m-1)}|
-
\frac12\E_q[\tau_h]Q_{ih}^{(m)}.
\]

Update $q(c_i)=\mathrm{Categorical}(r_{i1},\ldots,r_{iH})$:
\[
r_{ih}^{(m)}
\leftarrow
\frac{\exp(A_{ih}^{(m)})}
{\sum_{\ell=1}^H\exp(A_{i\ell}^{(m)})},
\qquad h=1,\ldots,H.
\]
}

Update $\delta$ (M-step):
\[
\delta^{(m)}
\leftarrow
\arg\max_{\delta>0}
\sum_{i=1}^N\sum_{h=1}^H
r_{ih}^{(m)}
\left[
-\frac12\log|\Omega_i(\delta)|
-\frac12\E_q[\tau_h]Q_{ih}(\delta)
\right].
\]

Calculate $\mathrm{ELBO}^{(m)}$ and
$\mathrm{ELBO}^{(m)}-\mathrm{ELBO}^{(m-1)}$\;
}

\caption{Variational EM Algorithm for Clustering Functional Data via Dirichlet Process Mixtures with Correlated Errors}
\label{alg:DPFunctionalOU}
\end{algorithm}

\begin{algorithm}[!ht]
\scriptsize
\DontPrintSemicolon

\KwData{
Observed functional data $\{Y_i,t_i,B_i\}_{i=1}^N$; truncation level $H$; hyperparameters $m_0$, $S_0$, $a_0$, $b_0$, and $\alpha$; total MCMC iterations $M$ and burn-in size $T$.
}

\KwResult{
Posterior samples of $c_i$, $v_h$, $\pi_h$, $\phi_h$, $\tau_h$, and $\delta$.
}

\textbf{Initialization}: initialize $c_i^{(0)}$, $v_h^{(0)}$, $\pi_h^{(0)}$, $\phi_h^{(0)}$, $\tau_h^{(0)}$, and $\delta^{(0)}$.\;

\For{$m=1,\ldots,M$}{

Construct
\[
\Omega_i^{(m-1)}=\Omega_i(\delta^{(m-1)}),
\qquad
\{\Omega_i^{(m-1)}\}_{rs}
=
\exp\{-\delta^{(m-1)}|t_{ir}-t_{is}|\}.
\]

\For{$h=1,\ldots,H$}{
Sample $\phi_h^{(m)}$ from
\[
\phi_h\mid - \sim N(\mu_h^\star,\Sigma_h^\star).
\]

Sample $\tau_h^{(m)}$ from
\[
\tau_h\mid - \sim \mathrm{Gamma}(a_h^\star,b_h^\star).
\]
}

\For{$i=1,\ldots,N$}{
Sample $c_i^{(m)}$ from
\[
P(c_i=h\mid-)
\propto
\pi_h^{(m-1)}
N\!\left(
Y_i;
B_i\phi_h^{(m)},
(\tau_h^{(m)})^{-1}\Omega_i^{(m-1)}
\right),
\qquad h=1,\ldots,H.
\]
}

\For{$h=1,\ldots,H-1$}{
Sample
\[
v_h^{(m)}
\sim
\mathrm{Beta}
\left(
1+n_h^{(m)},
\alpha+\sum_{\ell>h}n_\ell^{(m)}
\right),
\qquad
n_h^{(m)}=\sum_{i=1}^N I(c_i^{(m)}=h).
\]
}

Set $v_H^{(m)}=1$ and update
\[
\pi_h^{(m)}
=
v_h^{(m)}
\prod_{\ell<h}\left(1-v_\ell^{(m)}\right).
\]

Propose
\[
\log\delta^\star
=
\log\delta^{(m-1)}+\epsilon,
\qquad
\epsilon\sim N(0,s_\delta^2).
\]

Accept $\delta^\star$ with probability
\[
A=
\min\left\{
1,
\exp\left[
\ell(\delta^\star)-\ell(\delta^{(m-1)})
\right]
\frac{\delta^\star}{\delta^{(m-1)}}
\right\},
\]
where
\[
\ell(\delta)
=
-\frac12\sum_{i=1}^N\log|\Omega_i(\delta)|
-\frac12\sum_{i=1}^N
\tau_{c_i^{(m)}}^{(m)}
(Y_i-B_i\phi_{c_i^{(m)}}^{(m)})^\top
\Omega_i(\delta)^{-1}
(Y_i-B_i\phi_{c_i^{(m)}}^{(m)}).
\]

If accepted, set $\delta^{(m)}=\delta^\star$; otherwise set $\delta^{(m)}=\delta^{(m-1)}$.\;
}

Discard the first $T$ iterations and use the remaining samples for posterior inference.\;

\caption{MCMC Algorithm for Clustering Functional Data via Dirichlet Process Mixtures with OU Correlated Errors}
\label{alg:MCMC_DPFunctionalOU}
\end{algorithm}

\section{Simulation studies}\label{SimSec}

We conducted a series of simulation studies to evaluate the clustering performance, estimation accuracy, and computational efficiency of the proposed variational EM method. Particular attention was paid to the impact of within-curve dependence. Three simulation scenarios were considered, representing different levels of model complexity and different relationships between the data-generating mechanism and the working model. Within each scenario, four values of the OU decay parameter were considered,
\[
\delta \in \{3,5,8,12\},
\]
leading to Scenarios $j.1$--$j.4$, respectively, for $j=1,2,3$. Recall that under the OU correlation structure,
\[
\mathrm{Corr}\{Y_i(t),Y_i(s)\mid c_i=h\}
=
\exp\{-\delta|t-s|\}.
\]
Accordingly, smaller values of $\delta$ induce more persistent within-curve dependence, whereas larger values lead to a more rapid decay of correlation as the temporal separation increases.

For each simulation setting, 50 independent data sets were generated using random seeds $1,\ldots,50$. Unless otherwise specified, each functional observation was evaluated at $100$ equally spaced points over $[0,1]$, so that the spacing between adjacent observations was $1/99$. To provide a more interpretable characterization of the dependence induced by the four values of $\delta$, consider observations separated by ten grid intervals. Their correlation is approximately $0.7386$, $0.6035$, $0.4457$, and $0.2976$ for $\delta=3$, $5$, $8$, and $12$, respectively. Therefore, the four settings span a broad range of dependence strengths, from relatively persistent correlation across nearby observation times to substantially faster within-curve decorrelation.

The proposed method was fitted using cubic B-spline basis functions. The truncation level $H$ was chosen to be larger than the true number of clusters so that the effective number of clusters could be determined through the truncated Dirichlet process mixture rather than being fixed at its true value. The hyperparameters were set to $a_0=2$, $b_0=1$, and $\alpha=1$, with a diffuse Gaussian prior for the basis coefficients. 

\subsection{Simulation settings}

\paragraph{Scenario 1: data generated from the working B-spline model.}

The first scenario was designed to evaluate the proposed method under a correctly specified mean structure. We generated $K=3$ groups, with $50$ curves in each group. The cluster-specific mean functions were represented using six cubic B-spline basis functions,
\[
f_k(t)=\boldsymbol B(t)^\top\boldsymbol\phi_k,
\qquad k=1,2,3,
\]
where the coefficient vectors were
\[
\begin{aligned}
\boldsymbol\phi_1 &=
(1.5,1.0,1.6,1.8,1.0,1.5)^\top,\\
\boldsymbol\phi_2 &=
(1.8,0.6,0.4,2.6,2.8,1.6)^\top,\\
\boldsymbol\phi_3 &=
(1.2,1.8,2.2,0.8,0.6,1.8)^\top.
\end{aligned}
\]
For a curve belonging to group $k$, observations were generated according to
\[
\boldsymbol Y_i
=
\boldsymbol B\boldsymbol\phi_k+\boldsymbol\epsilon_i,
\]
where
\[
\boldsymbol\epsilon_i
\sim
N\left\{
\boldsymbol 0,
\sigma^2\boldsymbol\Omega(\delta)
\right\},
\qquad
\sigma=0.4,
\]
and
\[
\Omega_{rs}(\delta)
=
\exp\{-\delta|t_r-t_s|\}.
\]
The four decay parameters $\delta=3,5,8,$ and $12$ define Scenarios 1.1--1.4, respectively. The fitted model used six B-spline basis functions and a truncation level of $H=8$.

\paragraph{Scenario 2: misspecification of the basis representation.}

The second scenario was designed to examine robustness to misspecification of the functional basis. In contrast to Scenario 1, the data were generated using a Fourier basis but fitted using the proposed B-spline mixture model. Four groups were generated, with $50$ curves per group. Let
$\boldsymbol G(t)$ denote a Fourier basis containing five basis functions. The true mean functions were
\[
f_k(t)=\boldsymbol G(t)^\top\boldsymbol\phi_k^{(F)},
\qquad k=1,\ldots,4,
\]
where
\[
\begin{aligned}
\boldsymbol\phi_1^{(F)}
&=(1.40,-0.50,1.00,0.50,-0.80)^\top,\\
\boldsymbol\phi_2^{(F)}
&=(0.20,1.60,-0.70,1.20,0.60)^\top,\\
\boldsymbol\phi_3^{(F)}
&=(-1.00,0.80,1.70,-1.10,0.90)^\top,\\
\boldsymbol\phi_4^{(F)}
&=(1.20,-1.50,-0.40,1.80,-1.20)^\top.
\end{aligned}
\]
The observations were generated from
\[
\boldsymbol Y_i
=
\boldsymbol G\boldsymbol\phi_k^{(F)}
+
\boldsymbol\epsilon_i,
\]
with
\[
\boldsymbol\epsilon_i
\sim
N\left\{
\boldsymbol 0,
2^2\boldsymbol\Omega(\delta)
\right\}.
\]
The four values $\delta=3,5,8,$ and $12$ correspond to Scenarios 2.1--2.4. For all four settings, the proposed model was deliberately fitted using ten cubic B-spline basis functions, rather than the Fourier basis used for data generation. The truncation level was set to $H=8$. This setting therefore assesses whether the clustering procedure remains reliable when the working basis representation differs from the mechanism generating the underlying mean curves.

\paragraph{Scenario 3: nonlinear mean functions not generated from a basis expansion.}

The third scenario provides a more challenging setting in which the true mean functions were generated directly from nonlinear functions rather than from the B-spline representation assumed by the fitted model. Six groups were considered, each containing $50$ curves. For group $k$, $k=1, 2, ..., 6$, the mean function was
\[
f_k(t)
=
a_k+\cos(b_k\pi t)-t^2,
\]
where
\[
\boldsymbol a=(1.0,2.0,0.0,1.5,2.5,0.5)^\top
\]
and
\[
\boldsymbol b=(1.0,1.2,1.4,1.6,1.8,2.0)^\top.
\]
The observations were generated as
\[
Y_i(t_j)
=
a_k+\cos(b_k\pi t_j)-t_j^2+\epsilon_i(t_j),
\]
where
\[
\boldsymbol\epsilon_i
\sim
N\left\{
\boldsymbol 0,
0.4^2\boldsymbol\Omega(\delta)
\right\}.
\]
As in the previous scenarios, $\delta=3,5,8,$ and $12$ define Scenarios 3.1--3.4, respectively. The proposed method was fitted using eight cubic B-spline basis functions and a truncation level of $H=10$. This scenario simultaneously examines the ability of the proposed model to approximate nonlinear mean structures and to distinguish a larger number of relatively similar functional groups.

\subsection{Competing methods and performance measures}

We first compared the proposed variational EM approach, hereafter referred to as VBEM, with five competing methods. The first is the variational Bayes (VB) approach of \citet{Xian_2025}, which represents the cluster-specific mean functions using B-spline basis expansions and incorporates a curve-specific random intercept, while assuming conditionally independent residual errors. The remaining four competing methods are functional $K$-means \citep{TarpeyKinateder2003, Bande_2012}, fdaMocca \citep{ArnqvistEtAl2025}, funHDDC \citep{BouveyronJacques2011}, and SaS-Funclust \citep{Centofanti_2024}. These comparisons were conducted for all three scenarios and all four values of $\delta$. Whenever a competing method requires the number of clusters to be specified, it was supplied with the true number of groups so that the comparison focuses on clustering accuracy rather than selection of the number of clusters. In contrast, the proposed method was fitted with a truncation level larger than the true number of groups and was allowed to determine the effective number of occupied components.

Clustering performance was assessed using five complementary measures: clustering accuracy, V-measure \citep{Rosenberg_2007}, Rand index \citep{Rand_1971}, adjusted Rand index (ARI) \citep{HubertArabie_1985}, and Jaccard index \citep{Jaccard_1912}. Clustering accuracy was defined as the proportion of correctly classified curves after optimally matching the estimated cluster labels to the true group labels. Because numerical cluster labels are arbitrary, such label matching was performed only for the calculation of accuracy.

The V-measure evaluates clustering quality in terms of both homogeneity and completeness. Homogeneity is achieved when each estimated cluster contains observations predominantly from a single true group, whereas completeness is achieved when observations belonging to the same true group are assigned to the same estimated cluster. The V-measure is the harmonic mean of these two quantities and takes values between zero and one.

The remaining three measures are based on pairwise agreement. For any pair of curves, let TP denote the number of pairs placed in the same group under both the true and estimated partitions, TN the number placed in different groups under both partitions, FP the number placed in the same estimated cluster but different true groups, and FN the number belonging to the same true group but assigned to different estimated clusters. The Rand index is
\[
\mathrm{Rand}
=
\frac{\mathrm{TP}+\mathrm{TN}}
{\mathrm{TP}+\mathrm{TN}+\mathrm{FP}+\mathrm{FN}},
\]
while the Jaccard index is
\[
\mathrm{Jaccard}
=
\frac{\mathrm{TP}}
{\mathrm{TP}+\mathrm{FP}+\mathrm{FN}}.
\]
The adjusted Rand index further corrects the Rand index for agreement expected by chance. For all five measures, larger values indicate better agreement between the estimated and true partitions, with a value of one corresponding to perfect recovery.

For each combination of scenario, decay parameter, and clustering method, we report the mean and standard deviation of the five performance measures over the 50 simulated data sets.

\subsection{Comparison with MCMC}

In addition to the comparisons with existing functional clustering methods, we investigated the accuracy of the proposed variational approximation relative to an MCMC implementation of the same Dirichlet process mixture model. Because MCMC is substantially more computationally intensive, this comparison was performed for three representative settings rather than for every simulation configuration: Scenario 1.2 with $\delta=5$, Scenario 2.3 with $\delta=8$, and Scenario 3.4 with $\delta=12$. These settings represent, respectively, a correctly specified mean model, a basis-misspecified model, and the most complex nonlinear clustering setting considered in the study.

The MCMC and variational methods were compared from three perspectives. First, we compared their clustering performance using the same five measures described in the previous section. Second, we evaluated estimation of the OU decay parameter $\delta$. Across the 50 replications, the estimates of $\delta$ were summarized by their empirical mean, standard deviation, bias,
\[
\mathrm{Bias}(\hat\delta)
=
\frac{1}{R}
\sum_{r=1}^{R}
(\hat\delta_r-\delta_0),
\]
and mean squared error,
\[
\mathrm{MSE}(\hat\delta)
=
\frac{1}{R}
\sum_{r=1}^{R}
(\hat\delta_r-\delta_0)^2,
\]
where $R=50$ and $\delta_0$ denotes the true value.

Finally, we compared posterior inference for the cluster-specific mean functions. For the variational method,
\[
q(\boldsymbol\phi_h)
=
N(\boldsymbol\mu_h,\boldsymbol\Sigma_h),
\]
so the estimated mean curve is
\[
\widehat f_h(t)
=
\boldsymbol B(t)^\top\boldsymbol\mu_h,
\]
with pointwise posterior standard deviation
\[
\left\{
\boldsymbol B(t)^\top
\boldsymbol\Sigma_h
\boldsymbol B(t)
\right\}^{1/2}.
\]
The corresponding $95\%$ pointwise credible band was constructed as
\[
\widehat f_h(t)
\pm
1.96
\left\{
\boldsymbol B(t)^\top
\boldsymbol\Sigma_h
\boldsymbol B(t)
\right\}^{1/2}.
\]
For MCMC, a total of 5,000 iterations were generated, with the first
1,000 iterations discarded as burn-in. Posterior inference was therefore
based on the remaining 4,000 samples. Posterior samples of the basis coefficients were transformed into posterior samples of the mean function,
\[
f_h^{(s)}(t)
=
\boldsymbol B(t)^\top\boldsymbol\phi_h^{(s)},
\]
and the posterior mean and the $2.5\%$ and $97.5\%$ pointwise quantiles were used to construct the corresponding estimated curve and credible band. This comparison allows us to assess not only whether the variational approximation recovers similar point estimates to MCMC, but also whether it provides a comparable characterization of posterior uncertainty at a substantially lower computational cost.

\subsection{Simulation results}
\label{sec:simulation_results}
\paragraph{Comparison with existing functional clustering methods.}

Tables~\ref{tab:method_comparison_scenario1}--\ref{tab:method_comparison_scenario3}
summarize the clustering performance over the 50 replications. Across all three
scenarios, all four values of $\delta$, and all five evaluation criteria, VBEM
attains the highest average performance among the methods considered. This
consistent advantage is observed when the working mean model is correctly
specified, under basis misspecification, and when the true mean functions are
generated outside the assumed B-spline representation.

Under Scenario 1, where both the data-generating and fitted mean functions are
represented using B-splines, VBEM performs particularly well when the
within-curve dependence is strong. For Scenario 1.1 with $\delta=3$, VBEM
achieves an average accuracy of $0.9796$, compared with $0.8177$ for VB and
$0.7935$, $0.9235$, $0.8545$, and $0.9161$ for functional $K$-means,
fdaMocca, funHDDC, and SaS-Funclust, respectively. Similar differences are
observed for the other clustering criteria. In particular, the ARI of VBEM is
$0.9401$, compared with $0.6292$ for VB and values ranging from $0.5829$ to
$0.7994$ for the other competing methods. The corresponding Jaccard index is
$0.9234$, compared with $0.6111$ for VB and values between $0.5697$ and
$0.7707$ for the remaining competitors.

A notable pattern in Scenario 1 is that the difference between VBEM and VB
decreases as $\delta$ increases. The accuracy of VB increases from $0.8177$ at
$\delta=3$ to $0.8963$, $0.9196$, and $0.9424$ at $\delta=5$, $8$, and $12$,
respectively, while its ARI increases from $0.6292$ to $0.7307$, $0.7824$, and
$0.8398$. Under Scenario 1.4, VBEM still yields higher accuracy and ARI,
$0.9553$ and $0.8734$, respectively, but the difference is considerably smaller
than under Scenario 1.1. Functional $K$-means exhibits a similar improvement as
$\delta$ increases, attaining an accuracy of $0.9397$ and an ARI of $0.8331$
under Scenario 1.4. These results indicate that explicitly accounting for the
within-curve correlation structure is particularly beneficial when the dependence
is more persistent, whereas the advantage becomes smaller as the OU correlation
decays more rapidly.

Scenario 2 examines robustness to misspecification of the functional mean
representation, since the data are generated using a Fourier basis while the
fitted Bayesian models use B-splines. Despite this deliberate misspecification,
VBEM maintains high clustering accuracy across all four values of $\delta$.
The average accuracies for Scenarios 2.1--2.4 are $0.9851$, $0.9797$, $0.9703$,
and $0.9725$, respectively, while the corresponding ARI values are $0.9716$,
$0.9470$, $0.9237$, and $0.9289$. For Scenario 2.1, the Jaccard index of VBEM
is $0.9609$, compared with $0.7181$ for VB and $0.6710$, $0.7497$, $0.8416$,
and $0.8776$ for functional $K$-means, fdaMocca, funHDDC, and SaS-Funclust,
respectively.

The comparison between VBEM and VB again shows that the benefit of modeling the OU dependence is most evident for smaller values of $\delta$. The accuracy of VB increases from $0.8803$ under Scenario 2.1 to $0.9471$, $0.9549$, and $0.9644$ under Scenarios 2.2--2.4, respectively. The ARI obtained from VB similarly increases from $0.7717$ to $0.8658$, $0.8855$, and $0.9087$. Consequently, the difference
between VBEM and VB is substantial under Scenario 2.1 but much smaller under Scenario 2.4. Several other competing procedures display the same general pattern. For example, under Scenario 2.4, functional $K$-means attains an accuracy of $0.9653$ and an ARI of $0.9110$, while SaS-Funclust attains $0.9644$ and $0.9093$, respectively. Nevertheless, VBEM attains the highest mean value for each of the five clustering criteria in these settings. These results indicate that the proposed method maintains high clustering performance under the misspecification settings considered, while retaining the benefits of explicitly modeling within-curve dependence.

The advantage of VBEM is also evident under the nonlinear mean structures
considered in Scenario 3. When $\delta=3$, VBEM achieves an average accuracy of
$0.9627$, V-measure of $0.9151$, Rand index of $0.9763$, ARI of $0.9137$, and
Jaccard index of $0.8663$. The corresponding values for VB are $0.9328$,
$0.8587$, $0.9585$, $0.8491$, and $0.7772$, respectively. The average
accuracies for functional $K$-means, fdaMocca, funHDDC, and SaS-Funclust are
$0.9301$, $0.8819$, $0.9240$, and $0.9493$, respectively.

Clustering performance in Scenario 3 improves markedly as $\delta$ increases.
For VBEM, the average accuracy increases from $0.9627$ at $\delta=3$ to
$0.9743$, $0.9874$, and $0.9944$ at $\delta=5$, $8$, and $12$, respectively.
The corresponding ARI increases from $0.9137$ to $0.9401$, $0.9701$, and
$0.9866$, while the Jaccard index increases from $0.8663$ to $0.9051$,
$0.9515$, and $0.9779$. VB exhibits the same general trend, with its accuracy
increasing from $0.9328$ to $0.9615$, $0.9815$, and $0.9911$, and its ARI
increasing from $0.8491$ to $0.9111$, $0.9564$, and $0.9787$. The same narrowing of the performance gap at larger values of $\delta$ is also observed for functional $K$-means and SaS-Funclust. Under Scenario 3.4, functional $K$-means attains an accuracy of $0.9910$ and an ARI of $0.9786$,
while SaS-Funclust attains $0.9874$ and $0.9703$, respectively. VBEM achieves corresponding values of $0.9944$ and $0.9866$.

Taken together, the results across the 12 simulation settings demonstrate that
VBEM provides accurate and stable clustering under substantially different
data-generating mechanisms. The direct comparison with VB is especially
informative: although the random-intercept formulation can capture a common
curve-level shift, it does not explicitly represent the distance-dependent
within-curve correlation generated by the OU process. The improvement of VBEM over
VB is therefore most pronounced for smaller values of $\delta$, where the OU
dependence is stronger and more persistent, and systematically diminishes as
$\delta$ increases and the correlation decays more rapidly.

\begin{table}[p]
\centering
\caption{Clustering performance under Scenario 1. Entries are mean (SD) of performance metrics over 50 replications.}
\label{tab:method_comparison_scenario1}

\small
\setlength{\tabcolsep}{1.8pt}
\renewcommand{\arraystretch}{0.82}

\begin{tabular}{llccccc}
\toprule
Scenario & Method & Accuracy & V-measure & Rand & ARI & Jaccard \\
\midrule


\multirow{6}{*}{1.1}
& VBEM
& 0.9796 (0.0102)
& 0.9213 (0.0349)
& 0.9735 (0.0129)
& 0.9401 (0.0292)
& 0.9234 (0.0359) \\
& VB
& 0.8177 (0.1043)
& 0.6384 (0.1039)
& 0.8357 (0.0616)
& 0.6292 (0.1386)
& 0.6111 (0.1189) \\
& Functional $K$-means
& 0.7935 (0.0911)
& 0.6006 (0.0847)
& 0.8151 (0.0513)
& 0.5829 (0.1151)
& 0.5697 (0.0981) \\
& fdaMocca
& 0.9235 (0.0511)
& 0.7883 (0.0923)
& 0.9109 (0.0505)
& 0.7994 (0.1124)
& 0.7707 (0.1079) \\
& funHDDC
& 0.8545 (0.0488)
& 0.6888 (0.0616)
& 0.8435 (0.0431)
& 0.6545 (0.0882)
& 0.6336 (0.0728) \\
& SaS-Funclust
& 0.9161 (0.0371)
& 0.7678 (0.0653)
& 0.9026 (0.0351)
& 0.7800 (0.0790)
& 0.7469 (0.0783) \\

\midrule

\multirow{6}{*}{1.2}
& VBEM
& 0.9563 (0.0209)
& 0.8549 (0.0502)
& 0.9453 (0.0237)
& 0.8764 (0.0530)
& 0.8489 (0.0586) \\
& VB
& 0.8963 (0.0414)
& 0.7137 (0.0715)
& 0.8808 (0.0371)
& 0.7307 (0.0834)
& 0.6980 (0.0780) \\
& Functional $K$-means
& 0.8833 (0.0477)
& 0.6943 (0.0742)
& 0.8697 (0.0395)
& 0.7057 (0.0885)
& 0.6749 (0.0812) \\
& fdaMocca
& 0.8997 (0.0431)
& 0.7293 (0.0757)
& 0.8849 (0.0414)
& 0.7408 (0.0919)
& 0.7095 (0.0868) \\
& funHDDC
& 0.7923 (0.0819)
& 0.6158 (0.0884)
& 0.7991 (0.0665)
& 0.5674 (0.1176)
& 0.5681 (0.0784) \\
& SaS-Funclust
& 0.8627 (0.0568)
& 0.6790 (0.0810)
& 0.8550 (0.0470)
& 0.6740 (0.1036)
& 0.6482 (0.0929) \\

\midrule

\multirow{6}{*}{1.3}
& VBEM
& 0.9479 (0.0200)
& 0.8319 (0.0516)
& 0.9353 (0.0230)
& 0.8537 (0.0519)
& 0.8233 (0.0572) \\
& VB
& 0.9196 (0.0259)
& 0.7615 (0.0604)
& 0.9038 (0.0281)
& 0.7824 (0.0635)
& 0.7479 (0.0656) \\
& Functional $K$-means
& 0.9149 (0.0297)
& 0.7536 (0.0667)
& 0.8993 (0.0311)
& 0.7721 (0.0702)
& 0.7379 (0.0715) \\
& fdaMocca
& 0.8855 (0.0580)
& 0.7217 (0.0758)
& 0.8748 (0.0460)
& 0.7202 (0.0984)
& 0.6917 (0.0890) \\
& funHDDC
& 0.7578 (0.0834)
& 0.5957 (0.0895)
& 0.7820 (0.0669)
& 0.5336 (0.1183)
& 0.5439 (0.0866) \\
& SaS-Funclust
& 0.7971 (0.0753)
& 0.6139 (0.0827)
& 0.8143 (0.0471)
& 0.5835 (0.1035)
& 0.5709 (0.0898) \\

\midrule

\multirow{6}{*}{1.4}
& VBEM
& 0.9553 (0.0180)
& 0.8520 (0.0511)
& 0.9441 (0.0214)
& 0.8734 (0.0484)
& 0.8451 (0.0551) \\
& VB
& 0.9424 (0.0206)
& 0.8176 (0.0540)
& 0.9292 (0.0237)
& 0.8398 (0.0535)
& 0.8079 (0.0586) \\
& Functional $K$-means
& 0.9397 (0.0217)
& 0.8119 (0.0570)
& 0.9262 (0.0249)
& 0.8331 (0.0562)
& 0.8007 (0.0613) \\
& fdaMocca
& 0.9060 (0.0594)
& 0.7637 (0.0898)
& 0.8959 (0.0500)
& 0.7657 (0.1114)
& 0.7362 (0.1056) \\
& funHDDC
& 0.6947 (0.0565)
& 0.5572 (0.0841)
& 0.7566 (0.0377)
& 0.4754 (0.0841)
& 0.4982 (0.0622) \\
& SaS-Funclust
& 0.7861 (0.0556)
& 0.6032 (0.0572)
& 0.8048 (0.0302)
& 0.5617 (0.0678)
& 0.5501 (0.0553) \\

\bottomrule
\end{tabular}
\end{table}

\paragraph{Comparison with MCMC.}

We next assess the accuracy of the variational approximation by comparing VBEM with an MCMC implementation of the same Bayesian model. As shown in Table~\ref{tab:mcmc_clustering}, VBEM and MCMC yield nearly indistinguishable clustering results. Under Scenario 1.2, the average accuracies are $0.9563$ for VBEM and $0.9565$ for MCMC, while their ARI values are $0.8764$ and $0.8772$, respectively. Their V-measures are $0.8549$ and $0.8561$, and their Jaccard indices are $0.8489$ and $0.8498$. Thus, the differences between the two approaches are negligible relative to the Monte Carlo variability across simulated data sets.

The same level of agreement is observed under model misspecification. In Scenario 2.3, VBEM and MCMC achieve accuracies of $0.9703$ and $0.9702$, respectively, with corresponding ARI values of $0.9237$ and $0.9234$. Their Jaccard indices are also nearly identical, at $0.8922$ and $0.8920$. Under the more complex Scenario 3.4, both methods almost perfectly recover the true partition: the accuracies are $0.9944$ for VBEM and $0.9943$ for MCMC, and the ARI values are $0.9866$ and $0.9864$, respectively. Across the three selected scenarios and all five clustering measures, the largest absolute difference between the average performance of VBEM and MCMC is approximately $0.0013$.

The agreement between VBEM and MCMC extends to estimation of the OU decay parameter. Table~\ref{tab:mcmc_delta} reports the empirical mean, standard deviation, bias, and mean squared error of $\widehat{\delta}$. For Scenario 1.2, where the true value is $\delta=5$, the mean estimates are $4.8522$ for VBEM and $4.8252$ for MCMC. The corresponding biases are $-0.1478$ and $-0.1748$, with MSE values of $0.0934$ and $0.0988$, respectively. For Scenario 2.3 with $\delta=8$, the two procedures produce mean estimates of $8.0351$ and $8.0184$, with very similar standard deviations ($0.3504$ and $0.3507$) and MSE values ($0.1216$ and $0.1208$). Finally, for Scenario 3.4 with $\delta=12$, the mean estimates are essentially identical, at $11.8010$ for VBEM and $11.8009$ for MCMC. Their MSE values are $0.1254$ and $0.1362$, respectively.

\begin{figure*}[htbp]
\centering

\begin{minipage}{0.32\textwidth}
    \centering
    \textbf{Raw curves and true means}
\end{minipage}
\hfill
\begin{minipage}{0.32\textwidth}
    \centering
    \textbf{VBEM}
\end{minipage}
\hfill
\begin{minipage}{0.32\textwidth}
    \centering
    \textbf{MCMC}
\end{minipage}

\vspace{0.3em}


\begin{subfigure}[t]{0.32\textwidth}
    \centering
    \includegraphics[width=\linewidth]{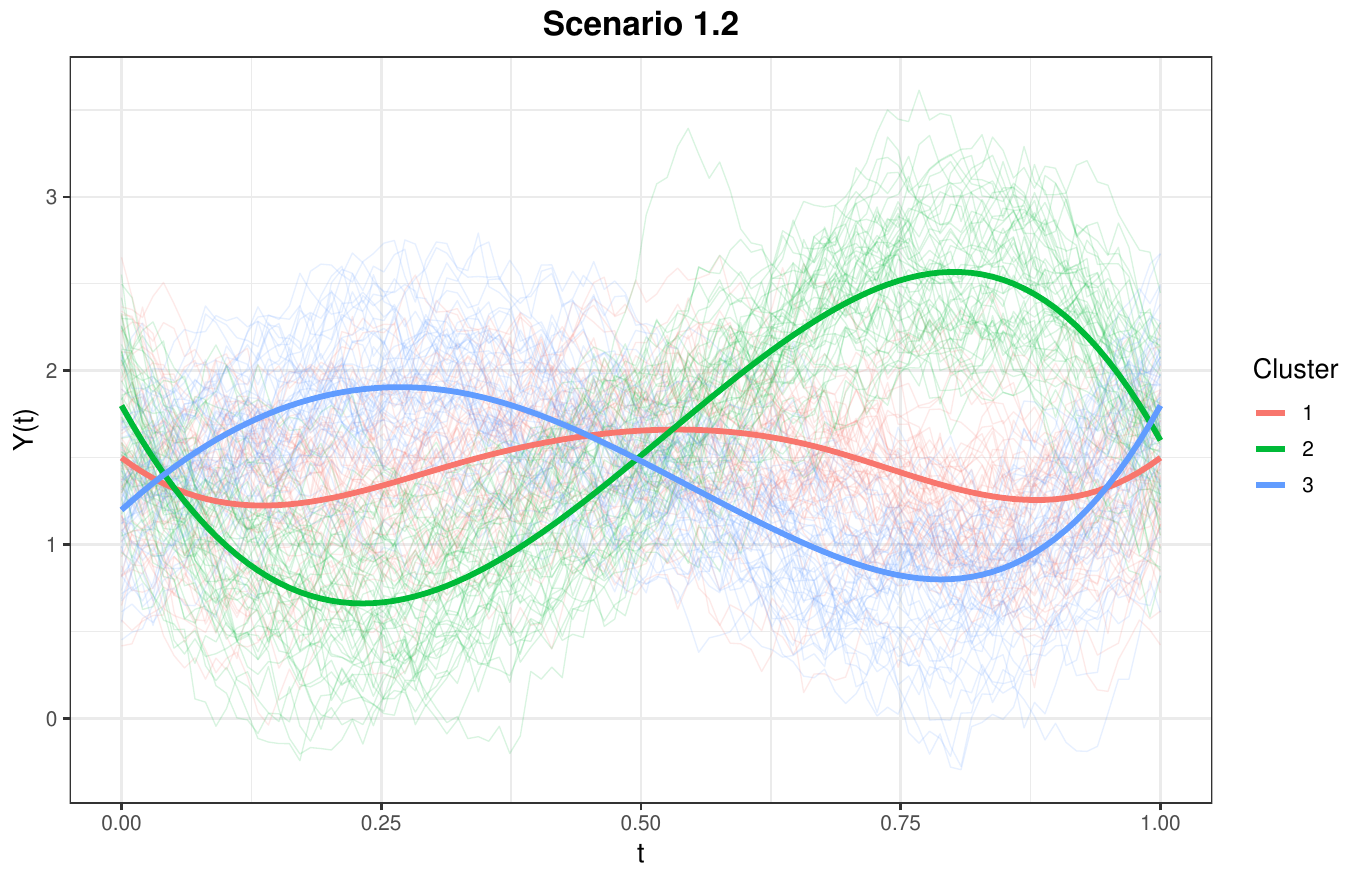}
    \caption{}
    \label{fig:sim_raw_12}
\end{subfigure}
\hfill
\begin{subfigure}[t]{0.32\textwidth}
    \centering
    \includegraphics[width=\linewidth]{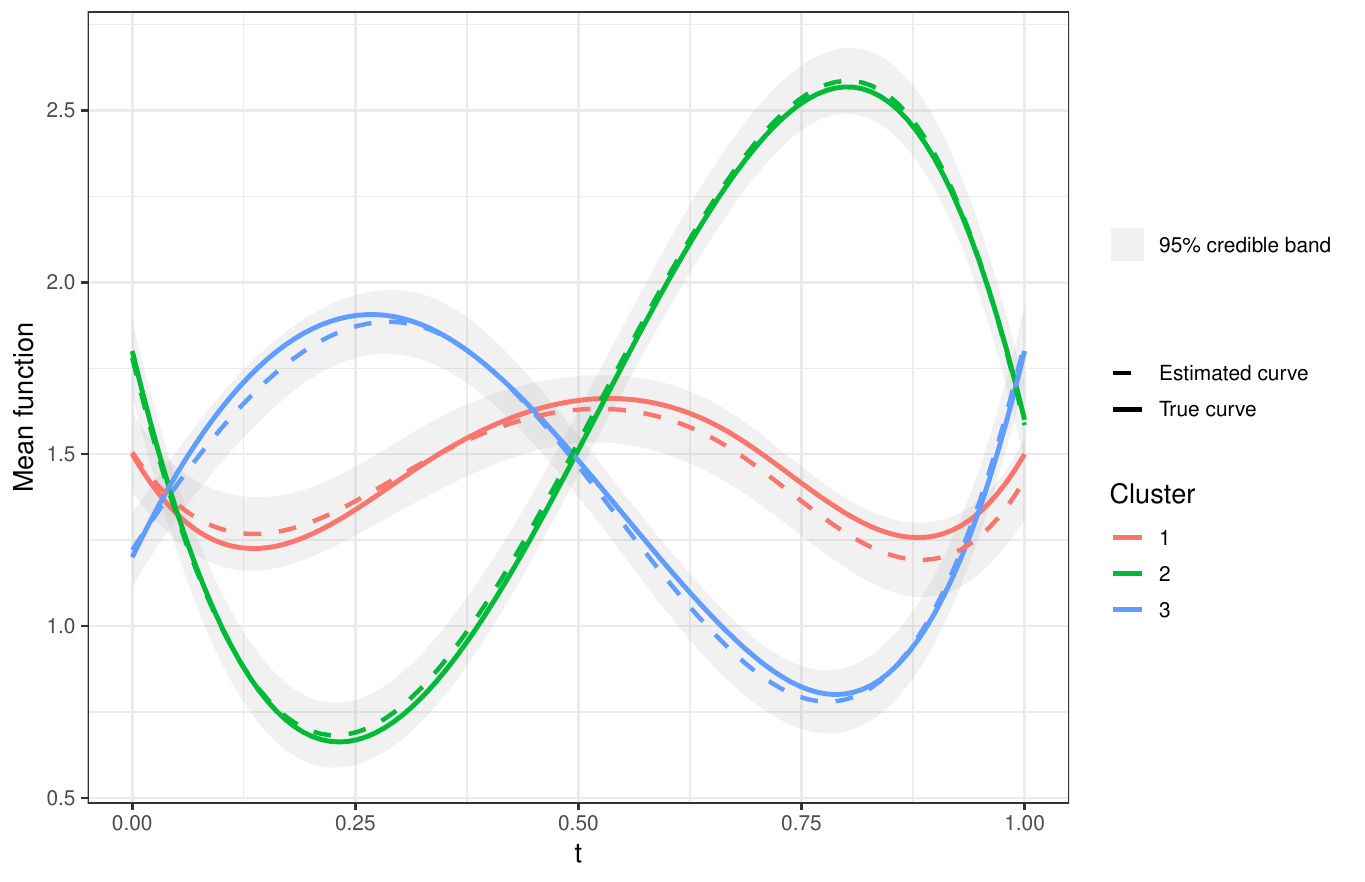}
    \caption{}
    \label{fig:sim_vbem_12}
\end{subfigure}
\hfill
\begin{subfigure}[t]{0.32\textwidth}
    \centering
    \includegraphics[width=\linewidth]{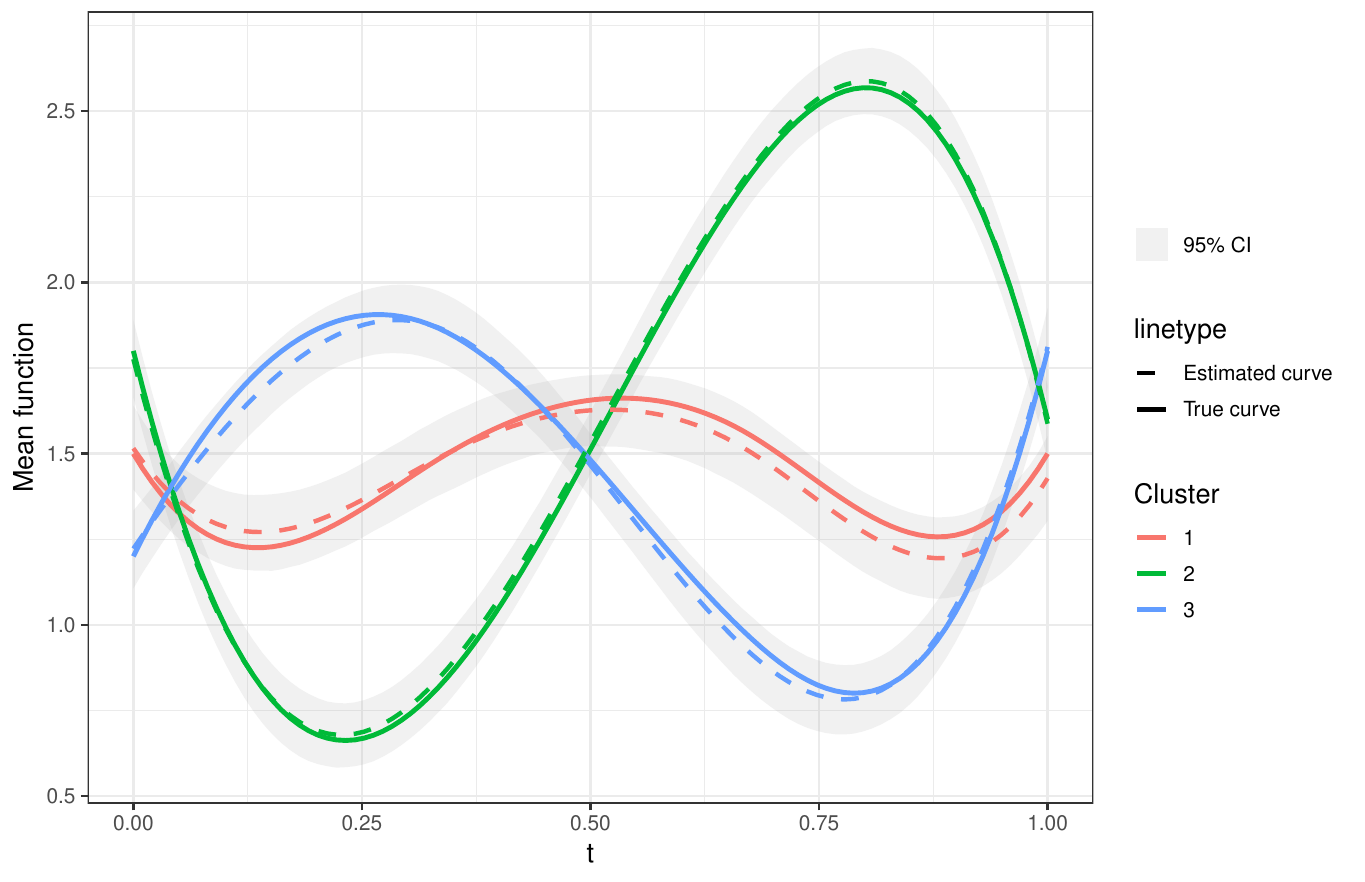}
    \caption{}
    \label{fig:sim_mcmc_12}
\end{subfigure}

\vspace{0.4em}


\begin{subfigure}[t]{0.32\textwidth}
    \centering
    \includegraphics[width=\linewidth]{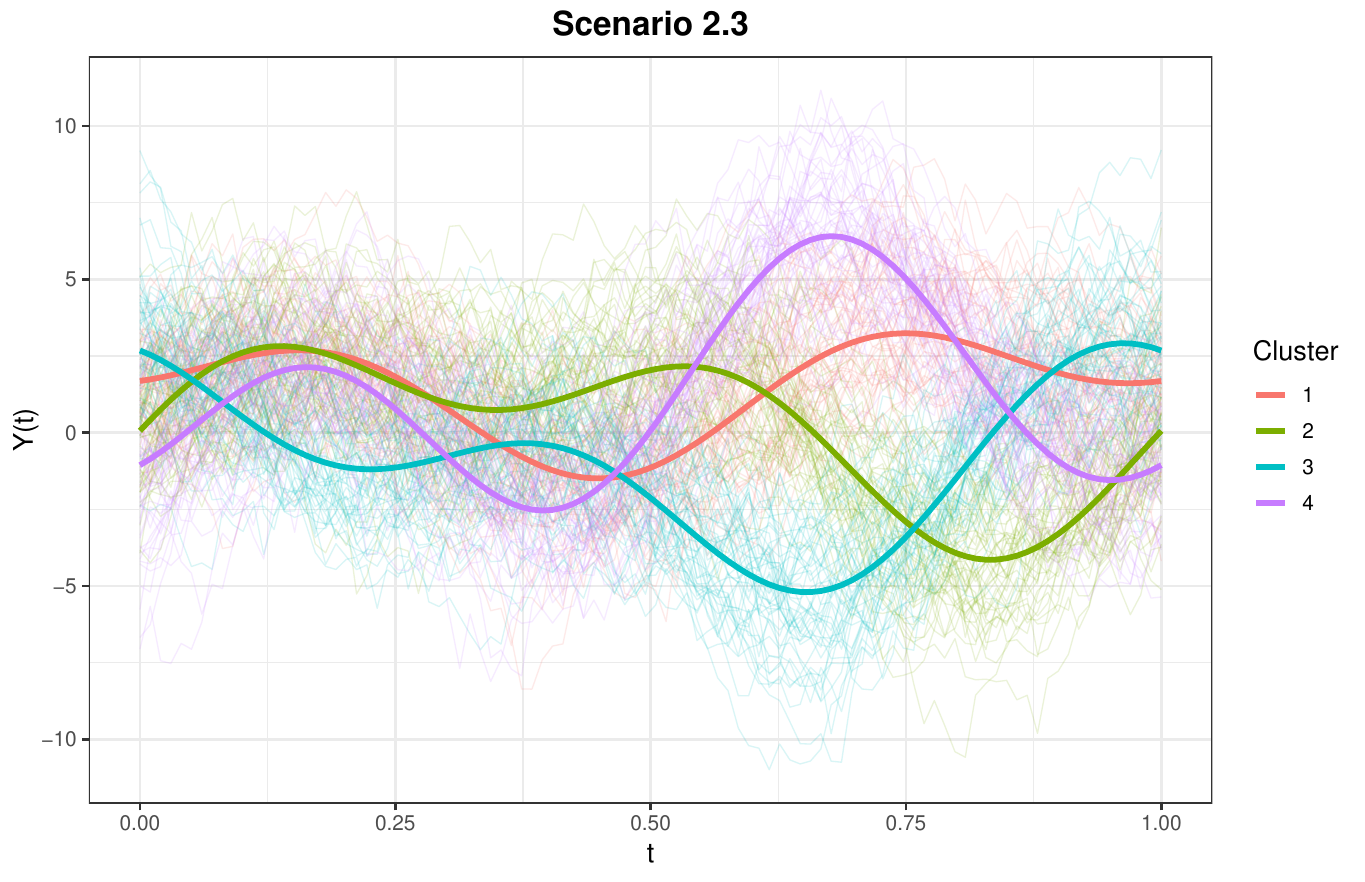}
    \caption{}
    \label{fig:sim_raw_23}
\end{subfigure}
\hfill
\begin{subfigure}[t]{0.32\textwidth}
    \centering
    \includegraphics[width=\linewidth]{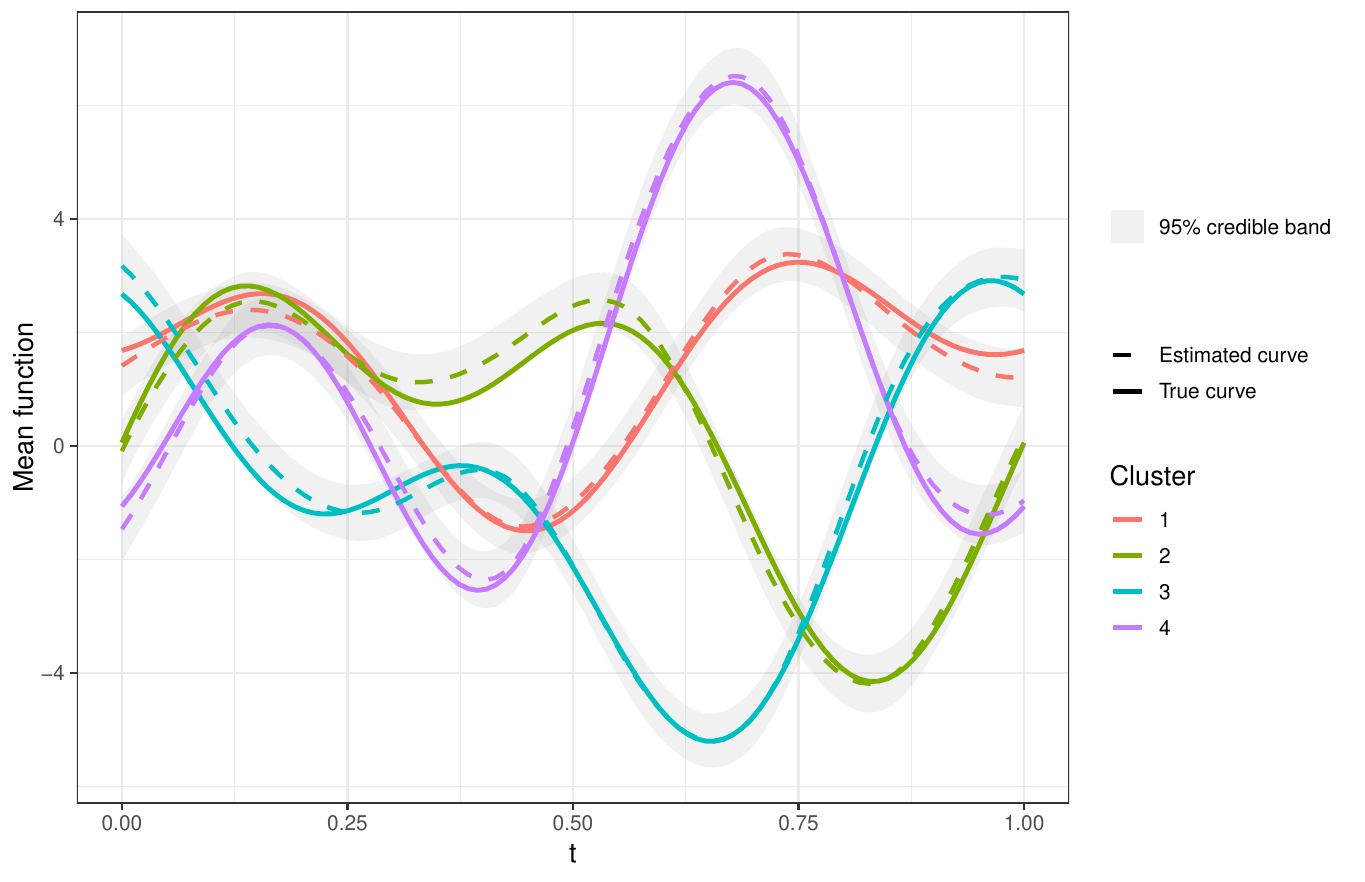}
    \caption{}
    \label{fig:sim_vbem_23}
\end{subfigure}
\hfill
\begin{subfigure}[t]{0.32\textwidth}
    \centering
    \includegraphics[width=\linewidth]{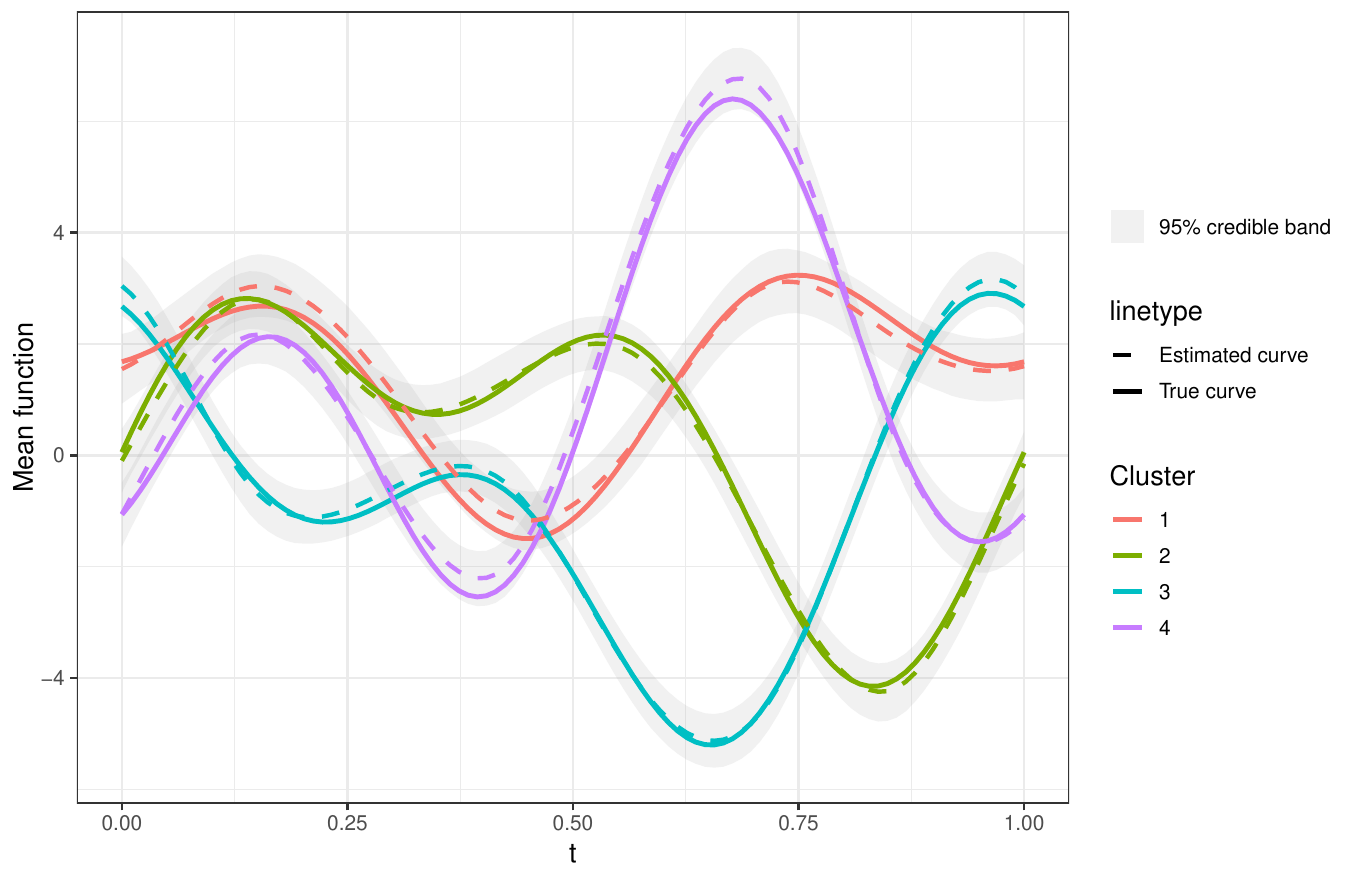}
    \caption{}
    \label{fig:sim_mcmc_23}
\end{subfigure}

\vspace{0.4em}


\begin{subfigure}[t]{0.32\textwidth}
    \centering
    \includegraphics[width=\linewidth]{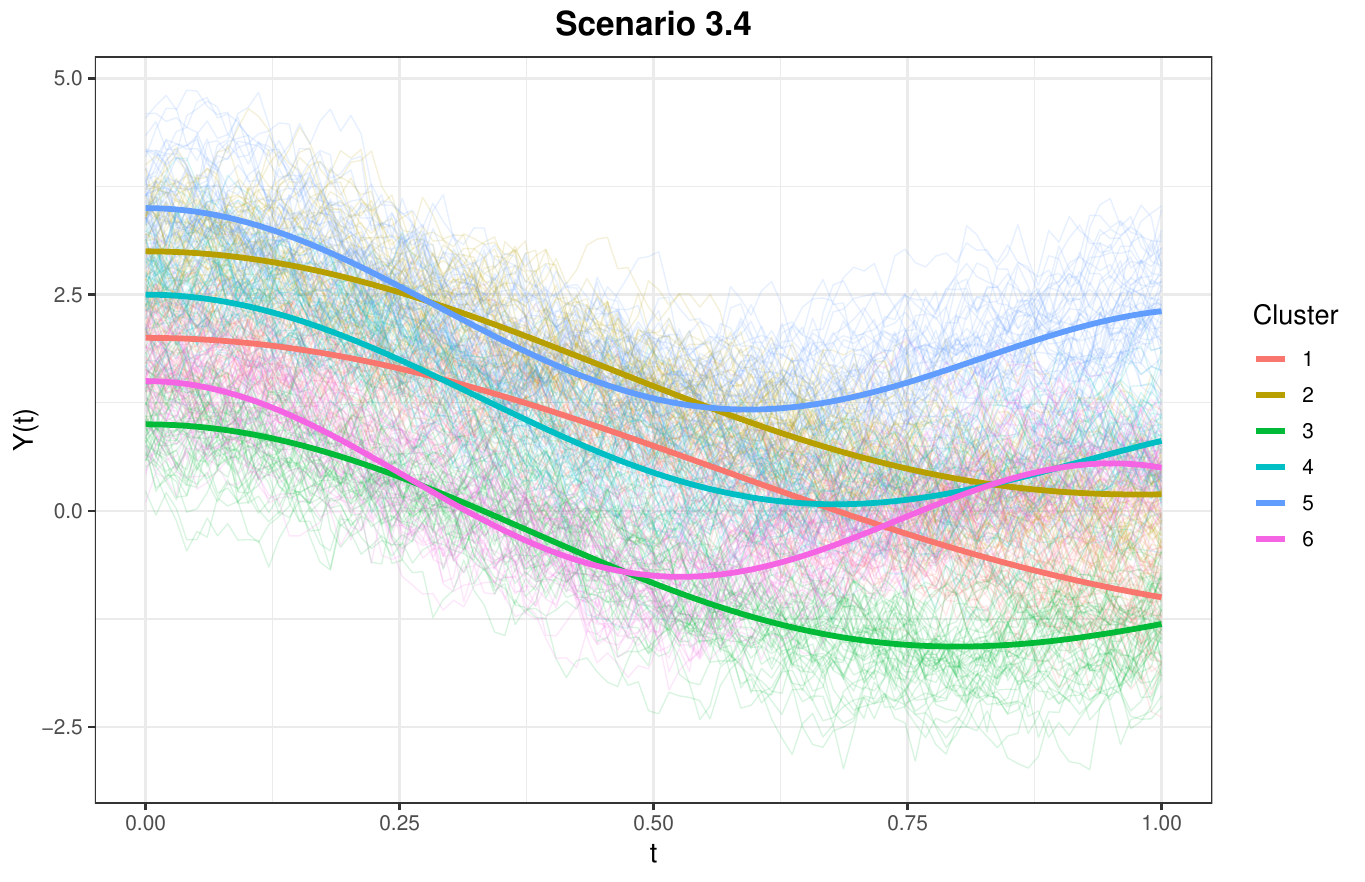}
    \caption{}
    \label{fig:sim_raw_34}
\end{subfigure}
\hfill
\begin{subfigure}[t]{0.32\textwidth}
    \centering
    \includegraphics[width=\linewidth]{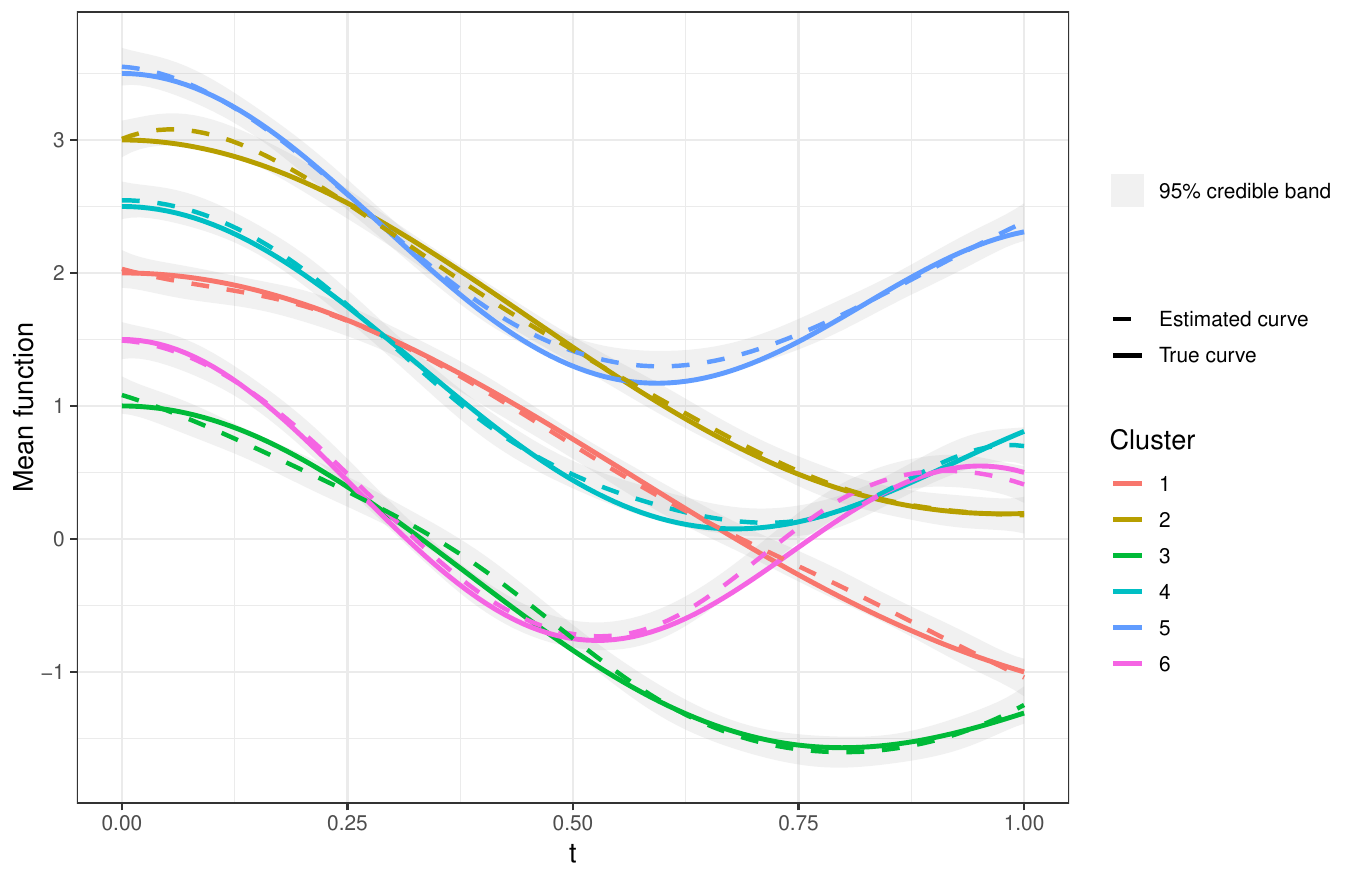}
    \caption{}
    \label{fig:sim_vbem_34}
\end{subfigure}
\hfill
\begin{subfigure}[t]{0.32\textwidth}
    \centering
    \includegraphics[width=\linewidth]{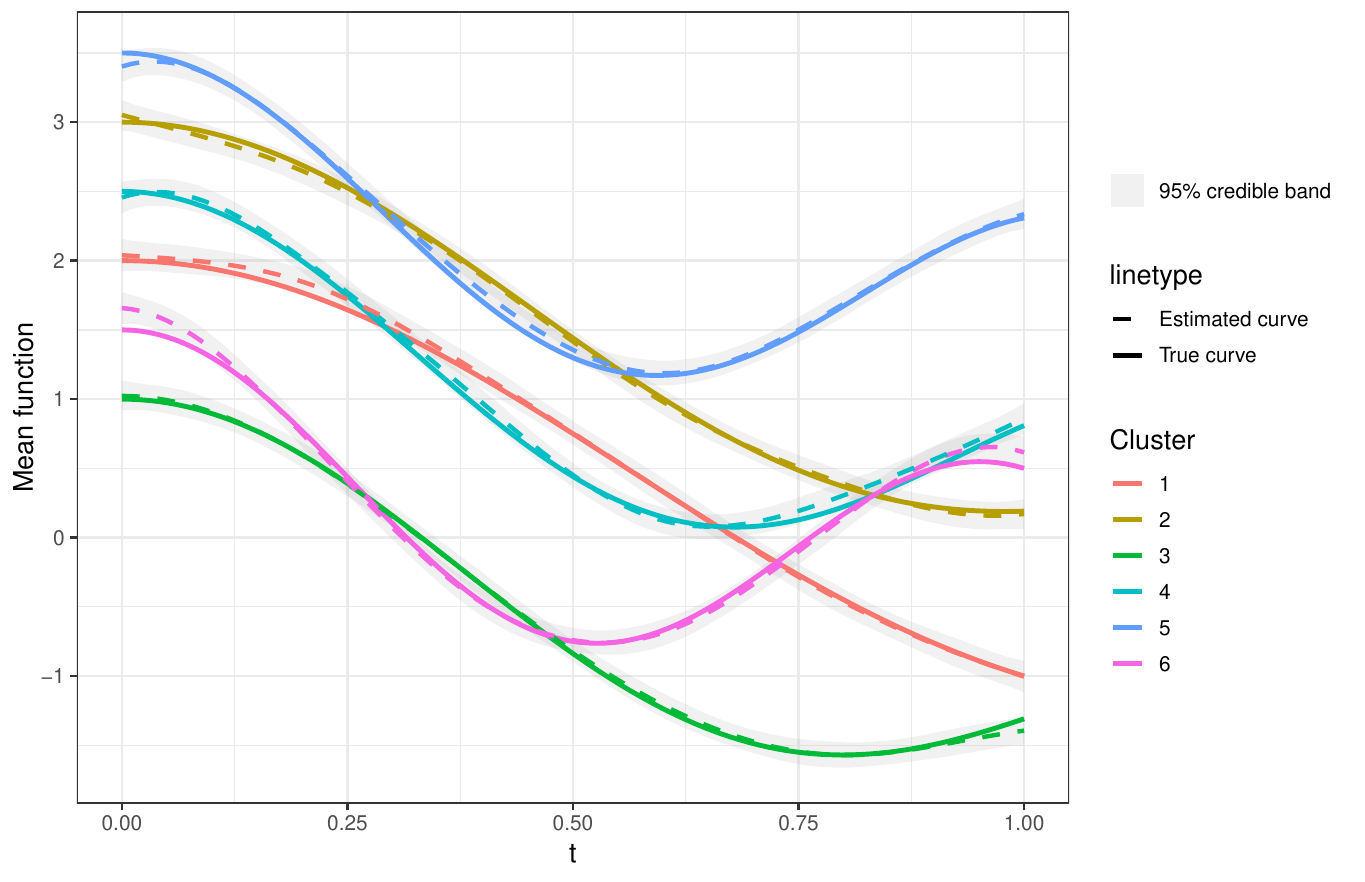}
    \caption{}
    \label{fig:sim_mcmc_34}
\end{subfigure}

\caption{
Simulated functional observations and estimation of the
cluster-specific mean functions. The rows correspond, from top to
bottom, to Scenarios 1.2, 2.3, and 3.4. The first column shows the
simulated functional observations together with the true mean
functions, while the second and third columns present the VBEM and
MCMC estimates, respectively. In the latter two columns, solid and
dashed curves denote the true and estimated mean functions,
respectively, and the shaded regions represent the corresponding
95\% credible bands.
}
\label{fig:simulation_curves}

\end{figure*}

Figure~\ref{fig:simulation_curves} provides a complementary visual
comparison of the recovery of the cluster-specific mean functions.
The three rows correspond to Scenarios 1.2, 2.3, and 3.4,
respectively, while the columns display the simulated functional
observations together with the true mean functions, the VBEM
estimates, and the MCMC estimates. Despite the substantial
within-cluster variation and overlap among the observed curves,
both VBEM and MCMC recover the underlying cluster-specific mean
structures well across all three settings.

Under Scenario 1.2, where the mean-function model is correctly
specified, the estimated curves from both methods closely track the
three true mean functions over the entire observation domain, and
the associated credible bands are highly consistent. The agreement
remains strong under Scenario 2.3, despite the deliberate mismatch
between the Fourier basis used to generate the data and the
B-spline basis used for model fitting. In particular, both methods
are able to recover the major peaks, troughs, and oscillatory
features of the four true mean functions even in the presence of
substantial noise and considerable overlap among the raw functional
observations. Scenario 3.4 provides a further challenge through six
nonlinear cluster-specific mean functions. Nevertheless, the VBEM
and MCMC estimates again follow the corresponding true curves
closely, with only minor local discrepancies.

\paragraph{Computational efficiency.}

The close statistical agreement between the proposed VBEM and MCMC is accompanied by a substantial reduction in computational cost. Table~\ref{tab:mcmc_time} reports the total running time for the 50 replications under the three selected scenarios. For Scenario 1.2, VBEM requires approximately $0.9483$ hours, whereas MCMC requires approximately $13.2920$ hours, corresponding to a speed-up of about $14.0$ times. Under Scenario 2.3, the total running time decreases from approximately $17.8218$ hours for MCMC to $0.8661$ hours for VBEM, representing a speed-up of approximately $20.6$ times.

The computational advantage becomes even more pronounced for Scenario 3.4, which involves six underlying clusters and a larger truncation level. In this setting, the 50 VBEM fits require approximately $1.1930$ hours in total, compared with approximately $27.0235$ hours for MCMC, corresponding to a speed-up of about $22.7$ times. Together with the nearly identical clustering performance and estimates of $\delta$, these results indicate that VBEM provides an effective approximation to the full MCMC analysis while reducing the computational burden by more than one order of magnitude. All algorithms were implemented in R version 4.5.1, and the simulations were conducted on a computer running macOS, equipped with a 4.05 GHz CPU and 8 GB of RAM.

\begin{table}[p]
\centering
\caption{Clustering performance under Scenario 2. Entries are mean (SD) of performance metrics over 50 replications.}
\label{tab:method_comparison_scenario2}

\small
\setlength{\tabcolsep}{1.8pt}
\renewcommand{\arraystretch}{0.82}

\begin{tabular}{llccccc}
\toprule
Scenario & Method & Accuracy & V-measure & Rand & ARI & Jaccard \\
\midrule


\multirow{6}{*}{2.1}
& VBEM
& 0.9851 (0.0488)
& 0.9702 (0.0320)
& 0.9891 (0.0219)
& 0.9716 (0.0528)
& 0.9609 (0.0606) \\
& VB
& 0.8803 (0.0903)
& 0.7855 (0.0928)
& 0.9147 (0.0481)
& 0.7717 (0.1276)
& 0.7181 (0.1377) \\
& Functional $K$-means
& 0.8606 (0.0851)
& 0.7421 (0.0890)
& 0.8990 (0.0441)
& 0.7297 (0.1171)
& 0.6710 (0.1233) \\
& fdaMocca
& 0.9084 (0.0499)
& 0.8330 (0.0650)
& 0.9269 (0.0335)
& 0.8045 (0.0889)
& 0.7497 (0.1010) \\
& funHDDC
& 0.9455 (0.0538)
& 0.8899 (0.0598)
& 0.9541 (0.0354)
& 0.8798 (0.0879)
& 0.8416 (0.1026) \\
& SaS-Funclust
& 0.9651 (0.0204)
& 0.9027 (0.0445)
& 0.9673 (0.0175)
& 0.9120 (0.0469)
& 0.8776 (0.0602) \\

\midrule

\multirow{6}{*}{2.2}
& VBEM
& 0.9797 (0.0107)
& 0.9378 (0.0299)
& 0.9803 (0.0101)
& 0.9470 (0.0273)
& 0.9239 (0.0379) \\
& VB
& 0.9471 (0.0186)
& 0.8501 (0.0447)
& 0.9501 (0.0165)
& 0.8658 (0.0444)
& 0.8180 (0.0546) \\
& Functional $K$-means
& 0.9356 (0.0239)
& 0.8234 (0.0516)
& 0.9403 (0.0201)
& 0.8393 (0.0539)
& 0.7864 (0.0637) \\
& fdaMocca
& 0.8864 (0.0462)
& 0.7886 (0.0568)
& 0.9106 (0.0283)
& 0.7604 (0.0753)
& 0.6987 (0.0823) \\
& funHDDC
& 0.9329 (0.0359)
& 0.8472 (0.0604)
& 0.9397 (0.0287)
& 0.8393 (0.0751)
& 0.7892 (0.0888) \\
& SaS-Funclust
& 0.9620 (0.0163)
& 0.8916 (0.0397)
& 0.9640 (0.0147)
& 0.9030 (0.0396)
& 0.8652 (0.0514) \\

\midrule

\multirow{6}{*}{2.3}
& VBEM
& 0.9703 (0.0130)
& 0.9140 (0.0326)
& 0.9716 (0.0119)
& 0.9237 (0.0321)
& 0.8922 (0.0433) \\
& VB
& 0.9549 (0.0178)
& 0.8725 (0.0443)
& 0.9575 (0.0159)
& 0.8855 (0.0429)
& 0.8428 (0.0553) \\
& Functional $K$-means
& 0.9541 (0.0196)
& 0.8704 (0.0461)
& 0.9568 (0.0170)
& 0.8839 (0.0458)
& 0.8409 (0.0583) \\
& fdaMocca
& 0.8752 (0.0416)
& 0.7667 (0.0526)
& 0.9016 (0.0257)
& 0.7363 (0.0683)
& 0.6721 (0.0738) \\
& funHDDC
& 0.9167 (0.0275)
& 0.8339 (0.0375)
& 0.9287 (0.0200)
& 0.8107 (0.0501)
& 0.7531 (0.0550) \\
& SaS-Funclust
& 0.9585 (0.0184)
& 0.8855 (0.0395)
& 0.9610 (0.0154)
& 0.8953 (0.0413)
& 0.8552 (0.0520) \\

\midrule

\multirow{6}{*}{2.4}
& VBEM
& 0.9725 (0.0120)
& 0.9184 (0.0300)
& 0.9736 (0.0111)
& 0.9289 (0.0298)
& 0.8991 (0.0400) \\
& VB
& 0.9644 (0.0144)
& 0.8965 (0.0347)
& 0.9661 (0.0131)
& 0.9087 (0.0353)
& 0.8723 (0.0463) \\
& Functional $K$-means
& 0.9653 (0.0142)
& 0.8988 (0.0352)
& 0.9669 (0.0129)
& 0.9110 (0.0347)
& 0.8753 (0.0457) \\
& fdaMocca
& 0.8746 (0.0512)
& 0.7802 (0.0576)
& 0.9050 (0.0291)
& 0.7461 (0.0770)
& 0.6837 (0.0826) \\
& funHDDC
& 0.9550 (0.0180)
& 0.8743 (0.0441)
& 0.9572 (0.0168)
& 0.8849 (0.0451)
& 0.8417 (0.0581) \\
& SaS-Funclust
& 0.9644 (0.0183)
& 0.8988 (0.0438)
& 0.9663 (0.0163)
& 0.9093 (0.0438)
& 0.8738 (0.0569) \\

\bottomrule
\end{tabular}

\end{table}

\begin{table}[p]
\centering
\caption{Clustering performance under Scenario 3. Entries are mean (SD) of performance metrics over 50 replications.}
\label{tab:method_comparison_scenario3}

\small
\setlength{\tabcolsep}{1.8pt}
\renewcommand{\arraystretch}{0.82}

\begin{tabular}{llccccc}
\toprule
Scenario & Method & Accuracy & V-measure & Rand & ARI & Jaccard \\
\midrule


\multirow{6}{*}{3.1}
& VBEM
& 0.9627 (0.0131)
& 0.9151 (0.0243)
& 0.9763 (0.0081)
& 0.9137 (0.0291)
& 0.8663 (0.0412) \\
& VB
& 0.9328 (0.0170)
& 0.8587 (0.0277)
& 0.9585 (0.0097)
& 0.8491 (0.0350)
& 0.7772 (0.0458) \\
& Functional $K$-means
& 0.9301 (0.0173)
& 0.8539 (0.0278)
& 0.9571 (0.0098)
& 0.8437 (0.0353)
& 0.7702 (0.0461) \\
& fdaMocca
& 0.8819 (0.0450)
& 0.7918 (0.0567)
& 0.9327 (0.0217)
& 0.7565 (0.0766)
& 0.6667 (0.0858) \\
& funHDDC
& 0.9240 (0.0243)
& 0.8504 (0.0361)
& 0.9533 (0.0146)
& 0.8312 (0.0515)
& 0.7555 (0.0650) \\
& SaS-Funclust
& 0.9493 (0.0153)
& 0.8878 (0.0283)
& 0.9683 (0.0091)
& 0.8847 (0.0330)
& 0.8253 (0.0455) \\

\midrule

\multirow{6}{*}{3.2}
& VBEM
& 0.9743 (0.0086)
& 0.9396 (0.0187)
& 0.9836 (0.0053)
& 0.9401 (0.0194)
& 0.9051 (0.0295) \\
& VB
& 0.9615 (0.0126)
& 0.9107 (0.0256)
& 0.9756 (0.0076)
& 0.9111 (0.0278)
& 0.8624 (0.0399) \\
& Functional $K$-means
& 0.9587 (0.0128)
& 0.9050 (0.0259)
& 0.9739 (0.0078)
& 0.9049 (0.0283)
& 0.8536 (0.0402) \\
& fdaMocca
& 0.9223 (0.0342)
& 0.8530 (0.0478)
& 0.9543 (0.0177)
& 0.8340 (0.0637)
& 0.7600 (0.0790) \\
& funHDDC
& 0.9209 (0.0299)
& 0.8470 (0.0350)
& 0.9523 (0.0161)
& 0.8283 (0.0535)
& 0.7521 (0.0645) \\
& SaS-Funclust
& 0.9662 (0.0103)
& 0.9231 (0.0207)
& 0.9786 (0.0062)
& 0.9220 (0.0227)
& 0.8782 (0.0336) \\

\midrule

\multirow{6}{*}{3.3}
& VBEM
& 0.9874 (0.0061)
& 0.9686 (0.0144)
& 0.9918 (0.0039)
& 0.9701 (0.0142)
& 0.9515 (0.0225) \\
& VB
& 0.9815 (0.0072)
& 0.9547 (0.0163)
& 0.9880 (0.0046)
& 0.9564 (0.0167)
& 0.9300 (0.0258) \\
& Functional $K$-means
& 0.9809 (0.0074)
& 0.9533 (0.0166)
& 0.9877 (0.0047)
& 0.9550 (0.0170)
& 0.9279 (0.0263) \\
& fdaMocca
& 0.9565 (0.0197)
& 0.9100 (0.0327)
& 0.9731 (0.0113)
& 0.9021 (0.0408)
& 0.8505 (0.0572) \\
& funHDDC
& 0.9284 (0.0235)
& 0.8550 (0.0347)
& 0.9562 (0.0134)
& 0.8414 (0.0467)
& 0.7681 (0.0590) \\
& SaS-Funclust
& 0.9793 (0.0087)
& 0.9513 (0.0180)
& 0.9867 (0.0055)
& 0.9515 (0.0200)
& 0.9226 (0.0308) \\

\midrule

\multirow{6}{*}{3.4}
& VBEM
& 0.9944 (0.0040)
& 0.9855 (0.0102)
& 0.9963 (0.0026)
& 0.9866 (0.0096)
& 0.9779 (0.0156) \\
& VB
& 0.9911 (0.0054)
& 0.9771 (0.0132)
& 0.9942 (0.0035)
& 0.9787 (0.0128)
& 0.9653 (0.0205) \\
& Functional $K$-means
& 0.9910 (0.0051)
& 0.9770 (0.0122)
& 0.9941 (0.0033)
& 0.9786 (0.0120)
& 0.9650 (0.0193) \\
& fdaMocca
& 0.9803 (0.0161)
& 0.9572 (0.0259)
& 0.9877 (0.0092)
& 0.9551 (0.0335)
& 0.9290 (0.0498) \\
& funHDDC
& 0.9217 (0.0287)
& 0.8478 (0.0355)
& 0.9524 (0.0161)
& 0.8288 (0.0539)
& 0.7528 (0.0657) \\
& SaS-Funclust
& 0.9874 (0.0071)
& 0.9693 (0.0155)
& 0.9919 (0.0045)
& 0.9703 (0.0163)
& 0.9519 (0.0257) \\

\bottomrule
\end{tabular}

\end{table}

\begin{table}[htbp]
\centering
\caption{Comparison of clustering performance between MCMC and the proposed VBEM. Entries are mean (SD) of performance metrics over 50 replications.}
\label{tab:mcmc_clustering}
\small
\setlength{\tabcolsep}{5pt}
\renewcommand{\arraystretch}{1.08}
\begin{tabular}{llccccc}
\toprule
Scenario & Method & Accuracy & V-measure & Rand & ARI & Jaccard \\
\midrule

\multirow{2}{*}{1.2}
& MCMC & 0.9565 (0.0210) & 0.8561 (0.0509) & 0.9457 (0.0237) & 0.8772 (0.0531) & 0.8498 (0.0588) \\
& VBEM & 0.9563 (0.0209) & 0.8549 (0.0502) & 0.9453 (0.0237) & 0.8764 (0.0530) & 0.8489 (0.0586) \\

\addlinespace[3pt]

\multirow{2}{*}{2.3}
& MCMC & 0.9702 (0.0147) & 0.9127 (0.0373) & 0.9715 (0.0134) & 0.9234 (0.0360) & 0.8920 (0.0480) \\
& VBEM & 0.9703 (0.0130) & 0.9140 (0.0326) & 0.9716 (0.0119) & 0.9237 (0.0321) & 0.8922 (0.0433) \\

\addlinespace[3pt]

\multirow{2}{*}{3.4}
& MCMC & 0.9943 (0.0039) & 0.9852 (0.0100) & 0.9963 (0.0025) & 0.9864 (0.0093) & 0.9777 (0.0151) \\
& VBEM & 0.9944 (0.0040) & 0.9855 (0.0102) & 0.9963 (0.0026) & 0.9866 (0.0096) & 0.9779 (0.0156) \\

\bottomrule
\end{tabular}%
\end{table}

\begin{table}[htbp]
\centering
\caption{Comparison of the estimation of the OU decay parameter $\delta$ between MCMC and the proposed VBEM over 50 replications.}
\label{tab:mcmc_delta}
\small
\setlength{\tabcolsep}{8pt}
\renewcommand{\arraystretch}{1.08}
\begin{tabular}{llcccc}
\toprule
Scenario & Method & Mean & SD & Bias & MSE \\
\midrule

\multirow{2}{*}{1.2}
& MCMC & 4.8252 & 0.2640 & -0.1748 & 0.0988 \\
& VBEM & 4.8522 & 0.2703 & -0.1478 & 0.0934 \\

\addlinespace[3pt]

\multirow{2}{*}{2.3}
& MCMC & 8.0184 & 0.3507 & 0.0184 & 0.1208 \\
& VBEM & 8.0351 & 0.3504 & 0.0351 & 0.1216 \\

\addlinespace[3pt]

\multirow{2}{*}{3.4}
& MCMC & 11.8009 & 0.3139 & -0.1991 & 0.1362 \\
& VBEM & 11.8010 & 0.2959 & -0.1990 & 0.1254 \\

\bottomrule
\end{tabular}
\end{table}

\begin{table}[htbp]
\centering
\caption{Computational cost of the proposed VBEM and MCMC for 50 replications.}
\label{tab:mcmc_time}
\small
\setlength{\tabcolsep}{10pt}
\renewcommand{\arraystretch}{1.08}

\begin{tabular}{lcc}
\toprule
Scenario & VBEM (hours) & MCMC (hours) \\
\midrule
1.2 & 0.9483 & 13.2920 \\
2.3 & 0.8661 & 17.8218 \\
3.4 & 1.1930 & 27.0235 \\
\bottomrule
\end{tabular}
\end{table}

\section{Application to real data}\label{AppSec}

To illustrate the proposed method, we applied it to the Canadian weather data \citep{RamsaySilverman2005}, available in the \texttt{fda} package. The data contain daily temperature and precipitation measurements from 35 weather stations across Canada, averaged over the period 1960--1994. Each station is therefore represented by a functional observation consisting of 365 daily temperature values over an annual cycle. Geographical coordinates are also available for all stations, allowing the resulting clusters to be examined from a spatial perspective.

\begin{figure}[htbp]
\centering
\includegraphics[width=0.78\textwidth]{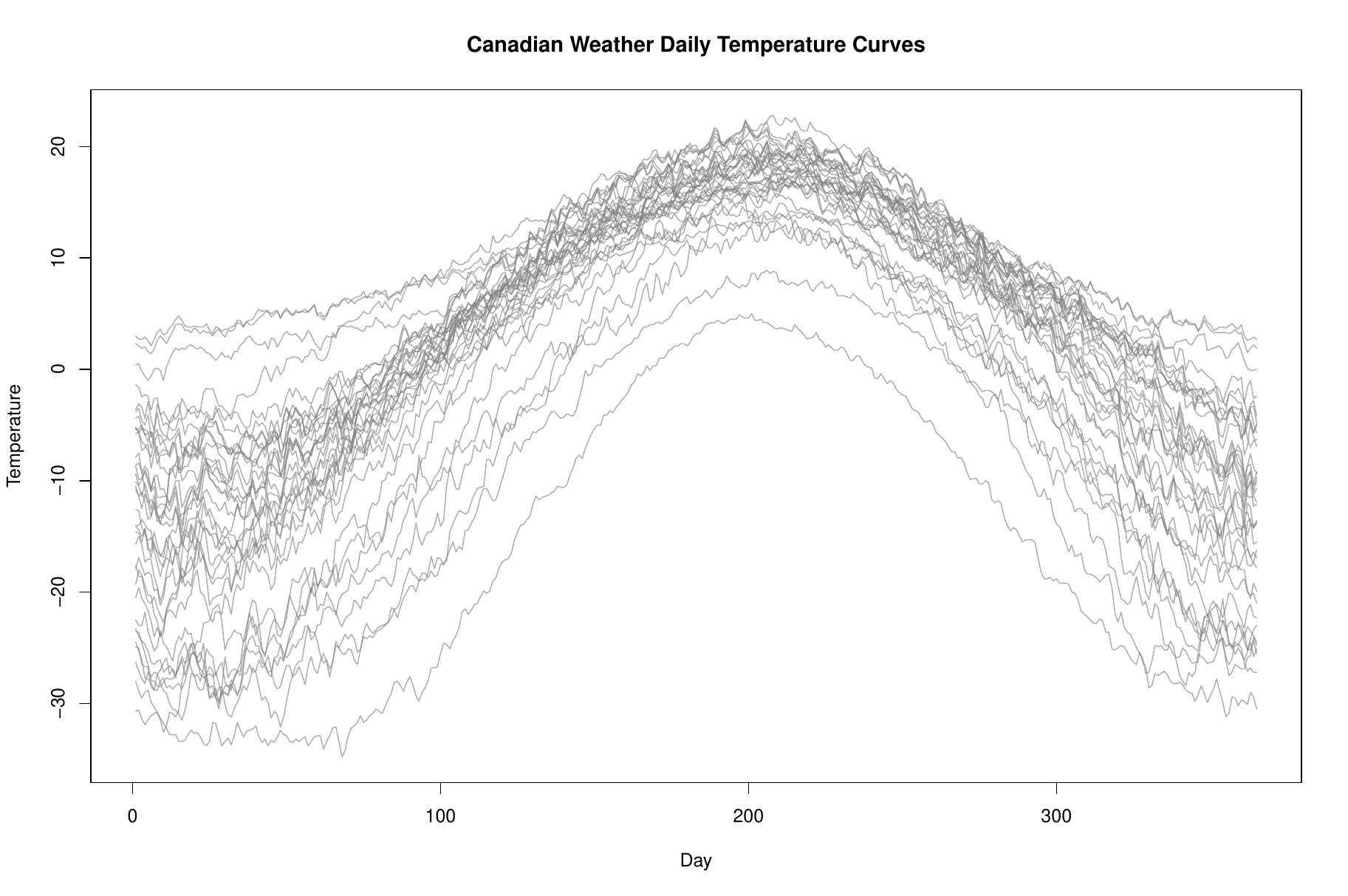}
\caption{
Daily average temperature curves for the 35 Canadian weather stations.
Each curve represents the average annual temperature trajectory of one
station over 365 days.
}
\label{fig:weather_raw}
\end{figure}

Figure~\ref{fig:weather_raw} displays the 35 observed temperature curves. Substantial heterogeneity is evident in both the overall temperature level and the amplitude of the annual cycle. In particular, the stations differ markedly in their winter temperatures, while the degree of separation is less pronounced during the warmer part of the year. These features suggest the presence of distinct annual temperature patterns and motivate a functional clustering analysis that simultaneously accommodates the pronounced serial dependence
within each trajectory.

For model fitting, the day index was rescaled to the unit interval $[0,1]$, and each cluster-specific mean function was represented using $M=6$ cubic B-spline basis functions. The truncation level of the Dirichlet process mixture was set to $H=8$, allowing the fitted model to contain more candidate mixture components than were expected to be substantively supported by the data. The hyperparameter settings were the same as those used in the simulation studies.

The variational parameters were initialized using an eight-cluster $K$-means partition of the observed curves. A small positive responsibility of $10^{-3}$ was assigned to the remaining components before row normalization. The initial stick-breaking parameters were subsequently obtained from the corresponding variational update equations. We initialized the OU decay parameter at $\delta^{(0)}=8$, and the variational EM algorithm was run with a maximum of 200 iterations and a convergence tolerance of $10^{-6}$. Following model fitting, a component was classified as active if its posterior effective cluster size,
\[
N_h^{\mathrm{eff}}=\sum_{i=1}^N r_{ih},
\]
exceeded 5. This criterion was used to distinguish substantively supported clusters from small residual components induced by the finite truncation.

\begin{figure}[htbp]
\centering

\begin{subfigure}[t]{0.47\textwidth}
    \centering
    \includegraphics[width=\linewidth]{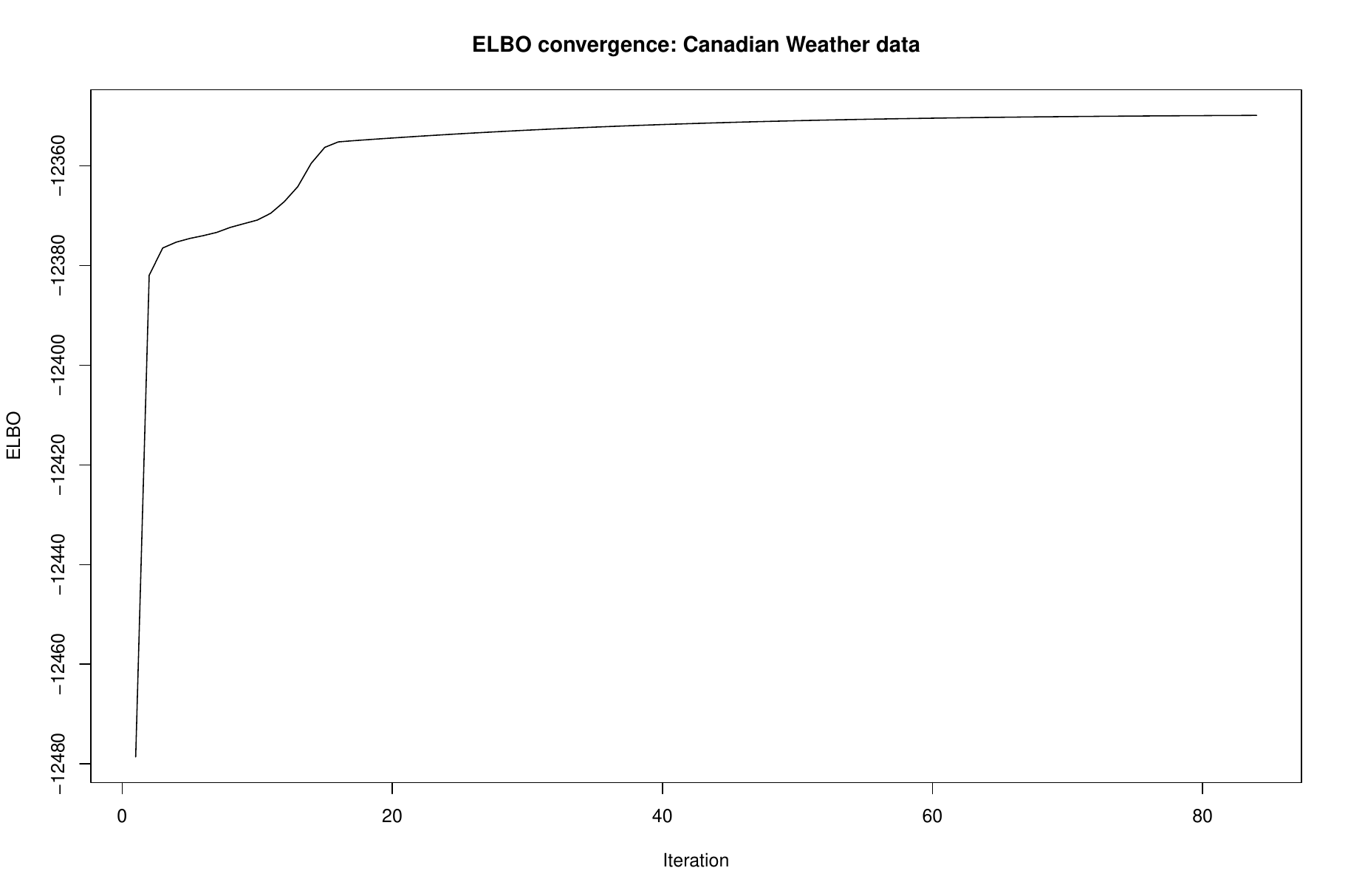}
    \caption{ELBO convergence.}
    \label{fig:weather_elbo}
\end{subfigure}
\hfill
\begin{subfigure}[t]{0.47\textwidth}
    \centering
    \includegraphics[width=\linewidth]{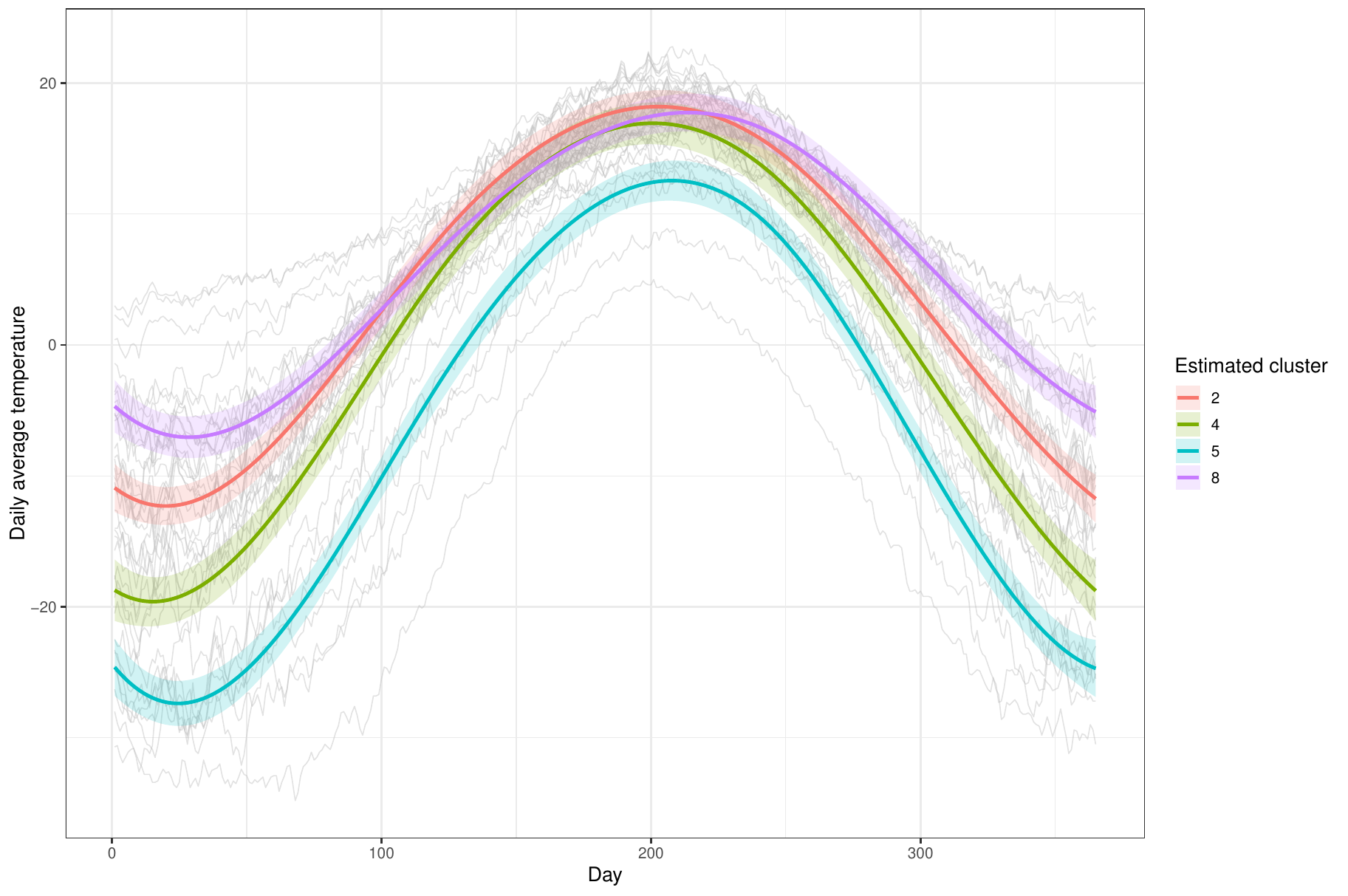}
    \caption{Estimated mean functions.}
    \label{fig:weather_estimated}
\end{subfigure}

\caption{
Model fitting and estimation results for the Canadian weather data.
Panel (a) shows the evolution of the evidence lower bound (ELBO) over
the variational EM iterations. Panel (b) shows the estimated
cluster-specific mean temperature functions for the four active
components (clusters), together with their 95\% pointwise credible bands; the
observed temperature curves are shown in gray.
}
\label{fig:weather_fit}

\end{figure}

The numerical behavior of the proposed algorithm was stable for the Canadian weather data set. As shown in Figure~\ref{fig:weather_elbo}, the ELBO increased rapidly during the initial iterations and subsequently approached a stable value, with the algorithm converging after 84 iterations. The final estimate of the OU decay parameter was $\widehat{\delta}=11.50$. Since the observation times were rescaled to $[0,1]$, a separation of $d$ days corresponds to a distance of $d/364$ on the rescaled time domain. Consequently, the fitted OU correlation between observations one week apart is
\[
\exp\left(-\widehat{\delta}\frac{7}{364}\right)=0.8016,
\]
while the estimated correlation between observations ten days apart is
\[
\exp\left(-\widehat{\delta}\frac{10}{364}\right)=0.7291.
\]
The estimated correlations therefore remain substantial even at weekly and ten-day separations, indicating pronounced within-curve dependence after accounting for the cluster-specific mean temperature trajectories.

Although the model was fitted with a truncation level of $H=8$, only four components, namely components 2, 4, 5, and 8, had posterior effective sizes greater than 5 and were therefore classified as active. The estimated cluster-specific mean functions for these components are displayed in Figure~\ref{fig:weather_estimated}, together with their 95\% pointwise credible bands. The four estimated trajectories reveal distinct annual temperature profiles, with particularly pronounced differences in winter temperature and in the amplitude of the seasonal cycle. The associated credible bands are relatively concentrated around the estimated mean trajectories, indicating that the dominant cluster-level seasonal patterns are estimated with reasonably high precision.

\begin{figure}[htbp]
\centering

\begin{subfigure}[t]{0.48\textwidth}
    \centering
    \includegraphics[width=\linewidth]{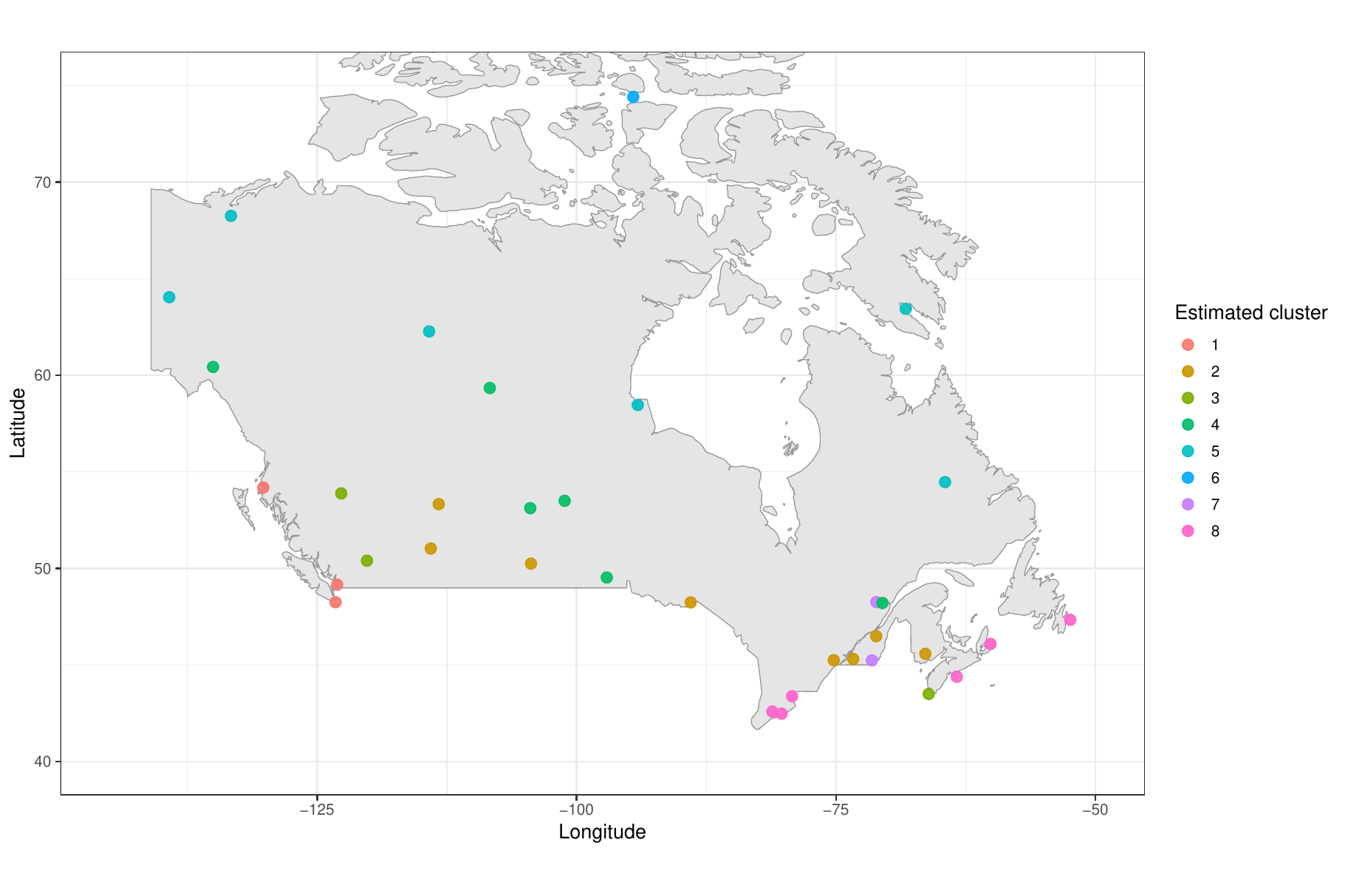}
    \caption{All estimated components.}
    \label{fig:weather_map_all}
\end{subfigure}
\hfill
\begin{subfigure}[t]{0.48\textwidth}
    \centering
    \includegraphics[width=\linewidth]{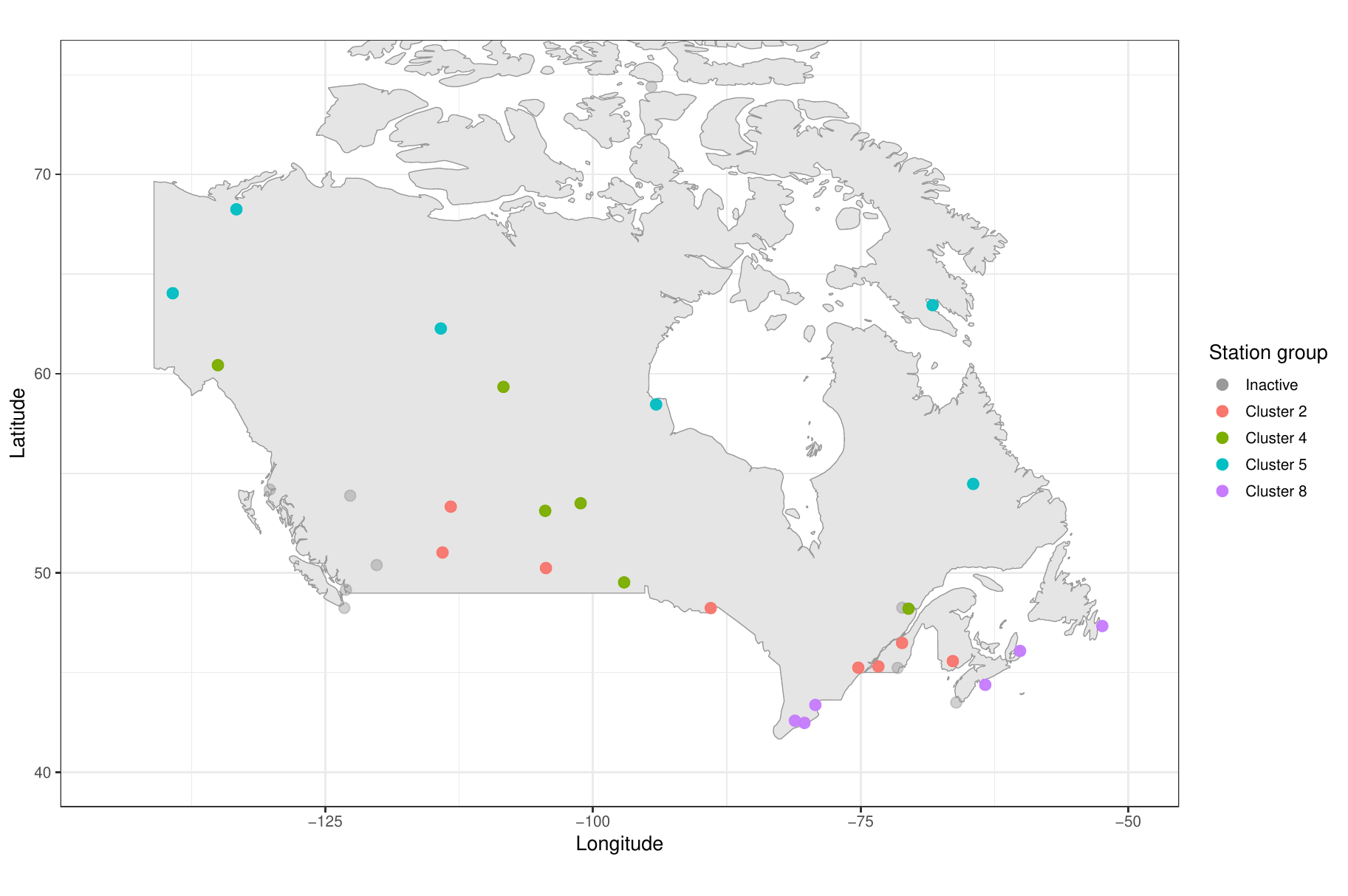}
    \caption{Active components.}
    \label{fig:weather_map_active}
\end{subfigure}

\caption{
Geographical distribution of the estimated clusters for the Canadian
weather stations. Panel (a) displays the assignments of all stations
to the eight components in the truncated Dirichlet process mixture.
Panel (b) highlights the four active components, defined as components
with posterior effective cluster size greater than 5; stations assigned
to inactive components are shown in gray.
}
\label{fig:weather_maps}

\end{figure}

The geographical distribution of the estimated clusters is presented in Figure~\ref{fig:weather_maps}. Figure~\ref{fig:weather_map_all} shows the assignments of all 35 weather stations to the eight components in the truncated mixture, including components supported by only a small number of stations. Figure~\ref{fig:weather_map_active} emphasizes the four active components, while stations assigned to inactive components are shown in gray. The active clustering exhibits a clear geographical structure. Stations assigned to the same active component tend to occupy geographically coherent regions, despite geographical information not being incorporated into the clustering model. This spatial organization provides additional qualitative evidence that the estimated groups capture meaningful differences in Canadian temperature regimes. At the same time, stations associated with the small inactive components appear as isolated observations rather than forming substantial geographical groups, supporting their interpretation as weakly supported residual components rather than major climatic clusters.

It is also noteworthy that the DIC-based analysis of \citet{Xian_2025} selected three clusters, whereas the proposed Dirichlet process model identified four substantively supported active components. This difference does not necessarily indicate conflicting clustering structures, but rather reflects the greater flexibility of the Dirichlet process formulation in allowing the data to support additional mixture components when warranted. The resulting four-cluster solution also exhibits interpretable differences in the estimated mean temperature trajectories and a clear geographical structure.

\section{Conclusion and Discussion}\label{ConDisSec}

In this paper, we developed a Bayesian model-based clustering framework for
functional data that simultaneously accommodates an unknown number of clusters
and within-curve dependence. Building upon the variational functional
clustering framework of \citet{Xian_2025}, the proposed model introduces two
major extensions. First, a truncated Dirichlet process mixture with a
stick-breaking representation is employed so that the effective number of
clusters can be learned from the data rather than specified in advance.
Second, an Ornstein--Uhlenbeck covariance structure is incorporated into the
within-curve error process to account explicitly for within-curve dependence among
repeated observations from the same functional unit. To achieve scalable
Bayesian inference, we developed a variational EM algorithm in which the
variational distributions of the latent allocation variables, stick-breaking
weights, cluster-specific basis coefficients, and precision parameters are
updated in the variational E-step, while the OU decay parameter is optimized
through the ELBO in the M-step.

The simulation studies demonstrated favorable performance of the proposed
method under a range of data-generating mechanisms and correlation strengths.
Across all 12 simulation settings and all five clustering criteria considered,
the proposed VBEM approach achieved the highest average clustering performance
among the methods examined. Its performance remained strong not only when the
working B-spline representation was correctly specified, but also when the
data were generated from a different basis system and when the true mean
functions followed nonlinear structures not generated from the fitted
B-spline model. These findings indicate that the procedure is reasonably robust
to misspecification of the functional mean representation. The comparison with
MCMC further showed that the variational approximation produced clustering
results and estimates of the correlation-decay parameter that were nearly
indistinguishable from those obtained using posterior sampling, while reducing
the computational cost by more than an order of magnitude. The close agreement
between the VBEM and MCMC estimates of the cluster-specific mean functions and
their credible bands provides additional support for the practical accuracy of
the variational approximation.

The analysis of the Canadian weather data further illustrated the usefulness
of the proposed approach for functional observations exhibiting substantial
serial dependence. Starting from an over-specified truncation level of
$H=8$, the fitted model identified four substantively supported temperature
patterns. The estimated mean functions revealed interpretable differences in
annual temperature profiles, particularly in winter temperature levels and
seasonal amplitudes, while the geographical distributions of the resulting
clusters exhibited clear spatial structure despite geographical information
not being incorporated into the model. Moreover, the estimated OU decay
parameter indicated substantial residual dependence between observations
separated by one week or ten days. These results illustrate
the ability of the proposed model to provide clustering, uncertainty
quantification, and interpretable characterization of within-curve dependence
within a unified probabilistic framework.

Several directions are worth pursuing in future work. First, the number of
B-spline basis functions is currently fixed in advance, and it would be useful
to develop a more adaptive procedure that allows the basis dimension to be
selected automatically. It would also be valuable to examine the proposed method on a broader range of
functional data applications and data structures. Finally, the methodology in
the present paper and that of \citet{Xian_2025} could be integrated into a
unified computational framework, covering different mixture formulations and
within-curve dependence structures. Developing an accompanying \textsf{R}
package would facilitate the practical use, comparison, and further extension
of these methods.








\begin{appendices}

\section{Derivations of update equations in the variational EM algorithm}\label{secA1}

In this appendix, we derive the coordinate ascent variational inference (CAVI) updates used in the variational E-step. Let $\Theta$ denote the collection of latent variables and model parameters. Under the mean-field factorization, the optimal variational factor for a parameter block $\theta_j$ is obtained from
\[
\log q(\theta_j)
=
\mathbb{E}_{-\theta_j}
\left[
\log p(Y,\Theta\mid\delta)
\right]
+\mathrm{const},
\]
where $\mathbb{E}_{-\theta_j}(\cdot)$ denotes expectation with respect to the variational distributions of all variables except $\theta_j$. By substituting the complete-data log-likelihood and retaining only those terms involving $\theta_j$, closed-form updates can be derived for the spline coefficients, precision parameters, cluster assignments, and stick-breaking weights.

\subsection{Derivation of the update for spline coefficients, $q(\phi_h)$}

The optimal variational factor for $\phi_h$ is obtained from
\[
\log q(\phi_h)
=
\mathbb{E}_{-\phi_h}
\left[
\log p(Y,\Theta\mid\delta)
\right]
+\mathrm{const}.
\]

From (\ref{eq:complete_loglik}), the terms involving $\phi_h$ are the likelihood and the Gaussian prior:
\[
\log p(Y,\Theta\mid\delta)
\propto
\sum_{i=1}^N
\mathbf{1}(c_i=h)
\log p(Y_i\mid c_i=h,\phi_h,\tau_h,\delta)
+
\log p(\phi_h).
\]

Taking expectation with respect to $q(c_i)$ and $q(\tau_h)$ yields
\begin{align*}
\log q(\phi_h)
\propto
&-\frac{1}{2}\mathbb{E}_q[\tau_h]
\sum_{i=1}^N r_{ih}
(Y_i-B_i\phi_h)^\top
\Omega_i(\delta)^{-1}
(Y_i-B_i\phi_h)
\\
&-\frac{1}{2}
(\phi_h-m_0)^\top
S_0^{-1}
(\phi_h-m_0).
\end{align*}

Expanding the quadratic terms and collecting those involving $\phi_h$, we obtain
\begin{align*}
\log q(\phi_h)
\propto
&-\frac{1}{2}
\phi_h^\top
\left[
S_0^{-1}
+
\mathbb{E}_q[\tau_h]
\sum_{i=1}^N
r_{ih}
B_i^\top
\Omega_i(\delta)^{-1}
B_i
\right]
\phi_h
\\
&+
\phi_h^\top
\left[
S_0^{-1}m_0
+
\mathbb{E}_q[\tau_h]
\sum_{i=1}^N
r_{ih}
B_i^\top
\Omega_i(\delta)^{-1}
Y_i
\right].
\end{align*}

Recognizing the kernel of a multivariate normal distribution, it follows that
\[
q(\phi_h)=\mathcal{N}(\mu_h,\Sigma_h),
\]
where
\[
\Sigma_h^{-1}
=
S_0^{-1}
+
\mathbb{E}_q[\tau_h]
\sum_{i=1}^N
r_{ih}
B_i^\top
\Omega_i(\delta)^{-1}
B_i,
\]
and
\[
\mu_h
=
\Sigma_h
\left[
S_0^{-1}m_0
+
\mathbb{E}_q[\tau_h]
\sum_{i=1}^N
r_{ih}
B_i^\top
\Omega_i(\delta)^{-1}
Y_i
\right].
\]

\subsection{Derivation of update for precision parameters, $q(\tau_h)$}

The optimal variational factor for $\tau_h$ is obtained from
\[
\log q(\tau_h)
=
\mathbb{E}_{-\tau_h}
\left[
\log p(Y,\Theta\mid\delta)
\right]
+\mathrm{const}.
\]
From the complete-data log-likelihood, the terms involving $\tau_h$ are the likelihood terms for observations assigned to component $h$ and the Gamma prior on $\tau_h$. Hence,
\begin{align*}
\log q(\tau_h)
\propto
&\;
\mathbb{E}_{-\tau_h}
\left[
\sum_{i=1}^N
\mathbf{1}(c_i=h)
\log p(Y_i\mid c_i=h,\phi_h,\tau_h,\delta)
+
\log p(\tau_h)
\right].
\end{align*}

Using
\[
Y_i\mid \{c_i=h,\phi_h,\tau_h,\delta\}
\sim
\mathcal{N}\left(B_i\phi_h,\tau_h^{-1}\Omega_i(\delta)\right),
\]
the terms depending on $\tau_h$ in the likelihood are
\[
\frac{n_i}{2}\log\tau_h
-
\frac{\tau_h}{2}
(Y_i-B_i\phi_h)^\top
\Omega_i(\delta)^{-1}
(Y_i-B_i\phi_h).
\]
Taking expectations with respect to $q(c_i)$ and $q(\phi_h)$ gives
\begin{align*}
\log q(\tau_h)
\propto
&\left[
\frac{1}{2}\sum_{i=1}^N r_{ih}n_i
\right]\log\tau_h
\\
&-
\frac{\tau_h}{2}
\sum_{i=1}^N r_{ih}
\mathbb{E}_q
\left[
(Y_i-B_i\phi_h)^\top
\Omega_i(\delta)^{-1}
(Y_i-B_i\phi_h)
\right] +
(a_0-1)\log\tau_h
-
b_0\tau_h .
\end{align*}

Collecting terms in $\log\tau_h$ and $\tau_h$, we obtain
\begin{align*}
\log q(\tau_h)
\propto
&\left[
a_0+\frac{1}{2}\sum_{i=1}^N r_{ih}n_i-1
\right]\log\tau_h
\\
&-
\left[
b_0+
\frac{1}{2}\sum_{i=1}^N r_{ih}
\mathbb{E}_q
\left[
(Y_i-B_i\phi_h)^\top
\Omega_i(\delta)^{-1}
(Y_i-B_i\phi_h)
\right]
\right]\tau_h .
\end{align*}

This is the kernel of a Gamma distribution under the shape--rate parameterization. Therefore,
\[
q(\tau_h)=\mathrm{Gamma}(\tilde a_h,\tilde b_h),
\]
where
\[
\tilde a_h
=
a_0+\frac{1}{2}\sum_{i=1}^N r_{ih}n_i,
\]
and
\[
\tilde b_h
=
b_0+
\frac{1}{2}\sum_{i=1}^N r_{ih}
\mathbb{E}_q
\left[
(Y_i-B_i\phi_h)^\top
\Omega_i(\delta)^{-1}
(Y_i-B_i\phi_h)
\right].
\]

\subsection{Derivation of update for cluster assignments, $q(c_i)$}

The optimal variational factor for $c_i$ is obtained from
\[
\log q(c_i)
=
\mathbb{E}_{-c_i}
\left[
\log p(Y,\Theta\mid\delta)
\right]
+\mathrm{const}.
\]
The terms involving $c_i$ are the likelihood contribution of $Y_i$ and the allocation probability:
\[
\log q(c_i)
\propto
\mathbb{E}_{-c_i}
\left[
\log p(Y_i\mid c_i,\phi,\tau,\delta)
+
\log p(c_i\mid v)
\right].
\]

For $c_i=h$, we have
\begin{align*}
\log q(c_i=h)
\propto
&\;
\mathbb{E}_q[\log \pi_h]
+
\frac{n_i}{2}\mathbb{E}_q[\log\tau_h]
-
\frac{1}{2}\log|\Omega_i(\delta)|
\\
&\quad
-
\frac{1}{2}\mathbb{E}_q[\tau_h]
\mathbb{E}_q
\left[
(Y_i-B_i\phi_h)^\top
\Omega_i(\delta)^{-1}
(Y_i-B_i\phi_h)
\right].
\end{align*}

Define
\[
\alpha_{ih}
=
\mathbb{E}_q[\log \pi_h]
+
\frac{n_i}{2}\mathbb{E}_q[\log\tau_h]
-
\frac{1}{2}\log|\Omega_i(\delta)|
-
\frac{1}{2}\mathbb{E}_q[\tau_h]
\mathbb{E}_q
\left[
(Y_i-B_i\phi_h)^\top
\Omega_i(\delta)^{-1}
(Y_i-B_i\phi_h)
\right].
\]
Since $c_i$ is a discrete latent variable taking values in $\{1,\ldots,H\}$, exponentiating and normalizing over $h$ gives
\[
q(c_i)=\mathrm{Categorical}(r_{i1},\ldots,r_{iH}),
\]
where
\[
r_{ih}
=
\frac{\exp(\alpha_{ih})}
{\sum_{m=1}^H \exp(\alpha_{im})}.
\]

\subsection{Derivation of update for stick-breaking weights, $q(v_h)$}

For $h=1,\ldots,H-1$, the optimal variational factor for $v_h$ is
obtained from
\[
\log q(v_h)
=
\mathbb{E}_{-v_h}
\left[
\log p(Y,\Theta\mid\delta)
\right]
+\mathrm{const}.
\]
The terms involving $v_h$ are the allocation model $p(c\mid v)$ and the prior $p(v_h)$. Therefore,
\[
\log q(v_h)
\propto
\mathbb{E}_{-v_h}[\log p(c\mid v)
+
\log p(v_h)].
\]

Under the stick-breaking representation,
\[
\pi_k=v_k\prod_{\ell<k}(1-v_\ell),
\]
and hence $v_h$ appears in $\pi_h$ through $\log v_h$ and in $\pi_k$ for all $k>h$ through $\log(1-v_h)$. Thus,
\[
\log p(c\mid v)
\stackrel{+}{=}
\left(\sum_{i=1}^N \mathbf{1}(c_i=h)\right)\log v_h
+
\left(\sum_{i=1}^N\sum_{\ell>h}\mathbf{1}(c_i=\ell)\right)\log(1-v_h),
\]
where $\stackrel{+}{=}$ denotes equality up to an additive constant.

Taking expectation with respect to $q(c)$ gives
\[
\mathbb{E}_{-v_h}[\log p(c\mid v)]
\stackrel{+}{=}
\left(\sum_{i=1}^N r_{ih}\right)\log v_h
+
\left(\sum_{i=1}^N\sum_{\ell>h}r_{i\ell}\right)\log(1-v_h).
\]

Since
\[
v_h\sim \mathrm{Beta}(1,\alpha),
\]
we have
\[
\log p(v_h)\stackrel{+}{=}(\alpha-1)\log(1-v_h).
\]
Combining these terms,
\[
\log q(v_h)
\stackrel{+}{=}
\left(\sum_{i=1}^N r_{ih}\right)\log v_h
+
\left(\alpha-1+\sum_{i=1}^N\sum_{\ell>h}r_{i\ell}\right)\log(1-v_h).
\]

This is the kernel of a Beta distribution. Therefore,
\[
q(v_h)=\mathrm{Beta}(\gamma_{h1},\gamma_{h2}),
\]
where
\[
\gamma_{h1}
=
1+\sum_{i=1}^N r_{ih},
\qquad
\gamma_{h2}
=
\alpha+\sum_{i=1}^N\sum_{\ell>h} r_{i\ell}.
\]
The final stick-breaking variable is fixed at $v_H=1$, so no
variational update is required for $v_H$.




\end{appendices}


\bibliography{sn-bibliography}

\end{document}